\documentclass{article}
\usepackage[utf8]{inputenc}
\usepackage[T1]{fontenc}
\usepackage{lmodern}
\usepackage{caption}
\usepackage{graphicx}
\usepackage[english]{babel}
\usepackage{afterpage}
\usepackage[authoryear]{natbib}
\usepackage{abstract}
\usepackage[obeyDraft, bordercolor=red, backgroundcolor=white]{todonotes}
\usepackage{amsmath}
\usepackage{amsthm}
\usepackage{bbm}
\usepackage{amssymb}
\usepackage{graphicx}
\usepackage{diagbox}
\usepackage[toc,page]{appendix}
\usepackage[hidelinks]{hyperref}
\usepackage{bm}
\usepackage{colortbl}
\usepackage{longtable}
\usepackage{adjustbox}
\usepackage{rotating}
\usepackage{lscape}
\usepackage[a4paper, left=1in, right=1in, top=1in, bottom=1in]{geometry}
\usepackage{setspace}
\usepackage{flafter}
\usepackage{morefloats}
\usepackage{array}
\usepackage{tabularx, booktabs}
\usepackage{authblk}
\usepackage{xr}
\usepackage{courier}
\usepackage[capposition=top]{floatrow}
\usepackage{multirow}
\usepackage{booktabs}
\usepackage{placeins}
\usepackage{slashbox}
\usepackage{endnotes}
\usepackage{etoolbox}
\usepackage{enumitem}

\newtheorem{assumption}{Assumption}
\newtheorem{definition}{Definition}
\newtheorem{proposition}{Proposition}

\newtheorem{remark}{Remark}
\newtheorem{corollary}{Corollary}

\DeclareMathOperator*{\argmin}{argmin}
\newcommand{\Lim}[1]{\raisebox{0.5ex}{\scalebox{0.8}{$\displaystyle \lim_{#1}\;$}}}

\newcolumntype{R}[2]{%
	>{\adjustbox{angle=#1,lap=\width-(#2)}\bgroup}%
	l%
	<{\egroup}%
}
\newcommand*\rot{\multicolumn{1}{R{30}{1em}}}

\newcolumntype{C}[1]{>{\centering\let\newline\\\arraybackslash\hspace{0pt}}m{#1}}

\newcommand\cites[1]{\citeauthor*{#1}'s\ (\citeyear{#1})}
\newcolumntype{s}{>{\hsize=.3\hsize}X}

\title{\textbf{Relaxing the Exclusion Restriction in Shift-Share Instrumental Variable Estimation}}

\author{Nicolas Apfel\thanks{Contact: n.apfel@surrey.ac.uk}}
\affil{School of Economics, University of Surrey}
\date{\today}
	
\begin{document}

\maketitle

\pagenumbering{gobble}

\begin{abstract}
Many economic studies use shift-share instruments to estimate causal effects. Often, all shares need to fulfil an exclusion restriction, making the identifying assumption strict. 
This paper proposes to use methods that relax the exclusion restriction by selecting invalid shares. 
I apply the methods in two empirical examples: the effect of immigration on wages and of Chinese import exposure on employment. In the first application, the coefficient becomes lower and often changes sign, but this is reconcilable with arguments made in the literature. In the second application, the findings are mostly robust to the use of the new methods.
\end{abstract}
\vspace{30px}
\newpage

\pagenumbering{arabic}

\section{Introduction}

The shift-share instrument is often used in applied economics to obtain estimates of causal effects. 
The numerous applications have spanned three decades, beginning with \cites{Bartik1991Who} seminal paper, and included several fields, such as migration, labor, international economics and many other. A shift-share strategy exploits shares at an earlier point in time and current aggregate-level changes to create an instrumental variable (IV). For example, the share of migrants from a certain origin country is interacted with the inflow from that country. The methods proposed in this paper are not restricted to the examples mentioned here, but can be applied to a wide range of studies in which shift-share instruments are used. 

The key assumptions and properties of shift-share instruments have been investigated only in very recent work. Centrally, in an important setting the exclusion restriction must hold for all initial shares, for the estimator to be consistent. In practice, a potentially large number of shares cannot have a direct effect on the outcome. This assumption is very strict, because it requires the researcher to have perfect structural knowledge about all shares, which is typically unavailable. The natural question to ask, therefore is: Is consistent estimation still possible when the exclusion restriction is violated for some but not all shares?

In this paper, my main contribution is to show how consistent shift-share estimation is possible, when not all shares fulfil the exclusion restriction. This paper is a practitioner's guide on how to select invalid shares in the shift-share setting, using two methods developed in statistical learning. I also extend one of the methods that I present to allow for multiple endogenous regressors. This was not possible so far and is a further contribution of the paper. 

The proposed methods go beyond existing econometric diagnostics typically applied in this setting. Rotemberg weights, proposed by \citet*{Goldsmith-Pinkham2020Bartik}, report the sensitivity to misspecification of the different shares. These weights often fail to provide clear-cut guidance as to which shares should finally be included in the construction of the instrument, because they tell the researcher how large the relative bias of the entire shift-share estimator stemming from the bias of a single industry is. Instead, with the methods proposed in this paper the researcher obtains an estimate for the identity of valid and invalid instruments and a consistent estimate, adjusted from the absolute bias. This bears the advantage that valid instruments do not need to be discarded just because their potential invalidity could lead to bias. 

Applying the new methods to the estimation of the effects of immigration and of Chinese import exposure on the US labor market illustrates that the shares selected as invalid in these applications are consistent with those discussed as problematic in the literature. So far, no way to locate invalid shares has been proposed. This paper fills this gap and provides a principled approach to share selection. To make the methods more widely accessible to practitioners, I also provide simple-to-use Stata-programs.

I begin by presenting the shift-share IV and its key identifying assumption, the exclusion restriction. 
In the shift-share approach, there are multiple class-specific shares and shifts which are interacted to produce the final instrument. 
\citet*{Goldsmith-Pinkham2020Bartik} show that if the exclusion restriction holds for each class-specific share, the IV estimator is consistent. That is, shares should not be directly correlated with the outcome variable through unobservable shocks or longterm effects. The exclusion restriction from the shares perspective is a sufficient condition for the consistency of the shift-share IV estimator.  
Instruments which fulfil the exclusion restriction are called \textit{valid}, while those that do not fulfil it are called \textit{invalid}. This definition of validity assumes that all instruments are related to the treatment. 
This exclusion restriction is very restrictive because it must hold for \textit{all} classes. While the general idea behind the shift-share IV is credible, typically structural relationships between instruments and outcome variables are difficult to exclude for each single class. 

In Section \ref{sec:modBartik}, I show how to obtain consistent estimators, when many shares are invalid. To achieve consistency, invalid shares are selected using two methods: the adaptive Least absolute shrinkage and selection operator (\textit{AL}) by \citet*{Windmeijer2019Use} and the Confidence Interval Method (\textit{CIM}) by  \citet*{Windmeijer2020Confidence}. The two methods have been developed primarily for the use in Mendelian randomization, which is the application of instrumental variable estimation in genetic epidemiology. In these applications, genetic markers are used as IVs when estimating the effect of an exposure on a health outcome. 

The key advantage of the leveraged methods is that they consistently select shares, which violate the exclusion restriction. When the \textit{majority} of instruments is valid, both metods have so-called \textit{oracle properties} and when the \textit{largest group} of shares fulfills the exclusion restriction the CIM has oracle properties. 
This means that asymptotically the post-selection estimators perform as well as if the researcher knew the identity of invalid IVs. 
Intuitively, these estimators both exploit the fact that just-identified estimates, which use only one valid IV at a time, converge against the same value. 
None of the existing methods allow for multiple endogenous regressors. I therefore propose a simple extension of the AL. Here, the exclusion restriction becomes stricter with increasing number of endogenous regressors.

To show the implications and generality of the presented methods in practice, in Sections \ref{sec:BP} and \ref{sec:ADH}, I apply them to two empirical examples. 
I apply the methods to the estimation of the effect of immigration on wages in the US as in \citet*{Altonji1991effects} and \citet*{Card2001Immigrant} and of Chinese import exposure on manufacturing employment, following \citet*{Autor2013china}. These two empirical examples are representative for a long series of applications in international and migration economics, which rely heavily on shift-share IVs. 

The use of the new methods suggests a lower effect of immigration on wages in the US. Using data from the US, the coefficients for the estimates that do not account for endogenous shares are positive. Using the estimators which adjust for shares selected as invalid, the estimates become smaller, often even change sign and retain statistical significance, when they were significant in the standard estimations. For example the effects on high-skilled wages change from 0.52 to -0.53 ($p < 0.01$). 
Among the selected there are many countries which were suspected of invalidity in the literature, such as the Philippines \citep*{Card2009Immigration}. When using a model with lagged immigration, the standard shift-share analysis produces estimates which point in different directions than expected. When using the proposed extension, the coefficients get switched in the expected direction again. Overall, the proposed methods seem to induce a large qualitative difference and should be used as a robustness check in migration settings.

When estimating the effect of import competition on employment, a large number of instruments is selected as invalid in some settings. 
It is noteworthy that many of the industry classes which have been discussed as problematic in \citet*{Autor2013china} and are likely to affect estimates \citep*{Goldsmith-Pinkham2020Bartik} are selected as invalid. 

These two applications illustrate the value of the proposed methods. 
The identity of chosen shares is consistent with economic intuition, because many of the origin countries and industries chosen as invalid are also discussed as being potentially problematic in the literature. This speaks for the plausibility of the outcomes of the new methods. Also, many shares which are similar to the discussed groups but were not specifically pointed out as being problematic in the literature, have been chosen by the new methods. Results can change qualitatively, when using the adjusted estimators. This shows that the methods are complementary to economic intuition and can help in designing an appropriate shift-share instrument. 

In an online Appendix I show an extension of the method to the multiple regressor case, a summary and illustrations of the methods as well as simulations confirming that in applications with shares as IVs with increasing sample size the performance approaches that of the estimator which uses only valid shares. This holds already for relatively small sample sizes. Second, allowing for weak instruments and stronger direct effects of the instruments on the outcome does not change the fact that the estimators converge to oracle performance quickly. In a third set of simulations, I test the performance of my extension to multiple endogenous regressors. The simulations confirm that with increasing number of regressors, the allowed number of invalid IVs becomes lower.

This paper relates to two strands of the literature. Recent work has brought forward two ways of motivating shift-share research designs: one which is justified by quasi-random shocks and one which stresses the exclusion restriction for shares only. 
\citet*{Borusyak2020Quasi} show consistency of the shift-share estimator when shocks are quasi-randomly assigned, conditionally on the shares. Also in this setting, \citet*{Adao2019Shift} discuss issues in inference. 
Unlike these two papers, \citet*{Goldsmith-Pinkham2020Bartik} put forward an interpretation of shift-share designs, according to which the shares express differential exposure to common shocks. The identification of the causal effect relies on the exclusion restriction for shares. Which setting is appropriate depends on the economic question at hand. 

This paper mainly relates to the exogenous share setting. Still, in situations in which it comes natural to think of the instrument from the perspective of random shifts, when many class-specific shifts are available these can be used to create multiple shift-share IVs or can be used as IVs directly. The selection methods then select among them instead of among the shares. For example, when imports are available for several high-income countries, in the \citet*{Autor2013china} example, the shifts can be used separately in a regression where the observations are industry-specific.  

The approach that I propose is helpful when both of these approaches fail. When the shifts are not random, consistency as derived by \citet*{Borusyak2020Quasi} does not follow. 
When the identifying assumption is motivated through the shares, but some of the class-specific shares are subject to criticism, but the general motivation of exogeneity still stands, the methods proposed here can offer interesting insights.

This paper also relates to the literature that proposes the use of machine learning methods for causal inference \citep*[e.g.][]{Athey2019Machine}. 
Shift-share IV estimation does not resemble a high-dimensional problem \textit{prima facie}, because there is only one instrument. Arguably however, all of the shares can be used as separate instruments and the need to select invalid shares substantially increases the complexity of the problem, making the use of machine learning methods appropriate. 
\citet*{Mullainathan2017Machine} have pointed out that in the context of economic research, machine learning methods lend themselves mostly to predictive tasks and less to causal inference. This paper provides a remedy for a commonly seen endogeneity problem, which threatens the reliability of causal inference in a wide range of economic studies.

\section{Shift-share instrumental variables and\\ the exclusion restriction}\label{sec:exclusion}

In this section, I present the shift-share setup and the exclusion restriction in terms of shares. I show under which conditions the exclusion restriction is fulfilled and show a setting in which it is plausible that some shares are valid and some invalid. A discussion of indications that a some valid - some invalid setting applies concludes this section.

\subsection{Endogeneity problem and instrument}

First, consider a linear model with a constant treatment effect $\beta$: 
\begin{equation}\label{eq:Model}
y_{lt} = \beta_0 + x_{lt} \beta + \theta_{lt} + u_{lt}\text{,}
\end{equation}
where $l$ indicates the location and $t$ the time period. A discussion of the constant treatment effect assumption can be found later in the text.  
The outcome variable is denoted by $y_{lt}$, $x_{lt}$ is the treatment, $u_{lt}$ is an idiosyncratic error term with $Cov(u_{lt}, x_{lt})=0$ and $\theta_{lt}$ denotes unobservable shocks which might be correlated with the treatment, i.e. $Cov(\theta_{lt}, x_{lt}) \neq 0$. 
For example, the outcome variable is employment growth in a certain region and year, the independent variable is growth of the immigrant share and the unobserved shocks $\theta_{lt}$ are labor demand shocks which might be correlated with the growth of the immigrant share. I abstract from covariates for ease of exposition.

In model \ref{eq:Model}, assume the treatment variable has the structure 
$$x_{lt} \equiv \sum_{j=1}^J z_{jlt} \cdot g_{jlt}\text{,}$$
where $j$ indicates a \textit{class} (e.g. the industry or the origin country of migrants), 
$z_{jlt}$ is the class-specific \textit{share} in a certain region and $g_{jlt}$ is the region-specific growth-rate (or \textit{shift}) of that class at time \textit{t}. For example, $z_{Mexico,CA,2020}$ is the share of Mexicans in California in 2020 and $g_{Mexico,CA,2020}$ is the inflow of migrants from Mexico to California in 2020. These shifts and shares are available for $J$ classes, i.e. origin countries, in the migration example.

In many settings, $x_{lt}$ can be subject to endogeneity problems such as correlation with unobserved shocks and reverse causality. In this model, the regressor is endogenous when $Cov(x_{lt}, \theta_{lt}) \neq 0$. In the migration context, Mexican migrants may have chosen to settle down in California precisely because of the high wages at destination. Part of the correlation that is measured with ordinary least-squares regressions would thus be due to migrant selection into regions. 

To circumvent this problem, a shift-share approach replaces components of the treatment variable by shares and shifts which are presumably unrelated with changes of the outcome variable. 
For example the share of Mexicans in California relative to Mexicans in the US is replaced with the same share, at a certain base period $t^0$ earlier in time (say 1990), while the growth rate of Mexican immigrants in California is replaced by its equivalent at the national level. The resulting shift-share IV is
\begin{equation}\label{eq:ShiftShareIV}
s_{lt} = \sum_{j=1}^J z_{jlt^0} \cdot g_{jt}\text{,}
\end{equation}
where $g_{jt}$ is the national growth rate of industry $j$ (i.e. the shift) at time $t$ and $s_{lt}$ is then used to instrument for $x_{lt}$. 

\subsection{Exclusion restriction}\label{sec:Exclusion}

The exclusion restriction is the key identifying assumption for any instrumental variable approach. In this setting, the exclusion restriction is stated in terms of shares. This is the setting proposed by  \citet*[GSS,][]{Goldsmith-Pinkham2020Bartik}. 
To show fulfilledness, violation and partial violation of the exclusion restriction, I set up a simple model. The structural equation is augmented by the shares $z_{jlt^0}$, with coefficients $\alpha_j$ which model the direct effects on the outcome. 
This is the definition of validity found in \citet*{Kang2016Instrumental}. 

The model becomes

\begin{align}
y_{lt} & = x_{lt} \beta + \sum\limits_{j=1}^J z_{jlt^0} \alpha_j + \theta_{lt} + u_{lt} \text{,} \label{eq:ModelAugmented} \\ 
x_{lt} & = s_{lt} \gamma_s + \varepsilon_{lt} \text{,} \label{eq:ModelFS} \\
x_{lt} & = \sum\limits_{j=1}^J z_{jlt^0} \gamma_j + \varepsilon_{lt} \text{,} \label{eq:ModelFS-Separate}
\end{align}
Equation \ref{eq:ModelFS} denotes the first stage. Relevance is given when $\gamma_s \neq 0$. When all shares are to be used as instruments, separately, relevance is given when $\gamma_j\neq0$ for all $j$ in equation \ref{eq:ModelFS-Separate}. This paper focuses on the exclusion restriction. Relevance is plausible because the underlying idea of this instrument is that immigrants settle in regions where they find communities of earlier migrants from their same country of origin, for example because they rejoin family members or there is a network of their country of origin which eases their arrival. This is why the shift-share instrument has also been called ``network'', ``enclave'' or ``past settlement instrument''. The higher probability to settle in regions in which communities of their same origin country can be found creates a correlation between past and present settlement, and the instrument is relevant. 

Shares might fail validity because they have a direct effect on the outcome, as measured by $\alpha_j$ but they might also need to be discarded because they are related to the outcome through unobservable shocks, $\theta_{lt}$. To show this, I allow for a non-zero correlation between current and past unobservable shocks:

\begin{equation}\label{eq:AR}
\theta_{lt} = \rho \theta_{lt^0} + \nu_{lt} \text{.}
\end{equation}
\noindent Now, assume that the past unobservable shocks can be written as 
\begin{equation}\label{eq:PastShocks}
\theta_{lt^0} = \sum\limits_{j=1}^J z_{jlt^0}\phi_j + \epsilon_{lt} \text{.}
\end{equation}
\noindent Then, the structural equation becomes

\begin{align}\label{eq:ModelAugmented-Again}
y_{lt} &= x_{lt} \beta + \sum\limits_{j=1}^J z_{jlt^0} \alpha_j + \rho (\sum\limits_{j=1}^J z_{jlt^0}\phi_j + \epsilon_{lt}) + \nu_{lt} + u_{lt} \\
&= x_{lt} \beta + \sum\limits_{j=1}^J z_{jlt^0} (\alpha_j + \rho \phi_j) + \xi_{lt} 
\end{align}
where $\xi_{lt} = \rho\epsilon_{lt} + \nu_{lt} + u_{lt}$. In order for the initial shares to be valid instruments, they should not be directly related with the outcome ($\alpha_j=0$) and they should not be related to the initial shocks ($\phi_j=0$) or there should be no serial correlation between initial and current shocks ($\rho=0$). 

Next, I summarize the above and introduce the definition of share validity in the context of this model to state the exclusion restriction more easily. 
\begin{definition}{\textbf{Validity}}\\
	A share $z_{jlt^0}$ is called valid when (1) $\alpha_j = 0$ and (2) at least one of the following two holds: $\rho=0$ or $\phi_j=0$.
\end{definition}
\noindent In the migration example, the first part of validity means that there is no adjustment through other factors of production. The second part means that unobserved shocks are not related to initial shares and/or these shocks are not correlated over time. 
The strict exclusion restriction can now be stated as
\begin{assumption}\label{ass:excls}
	\textbf{Strict exclusion restriction: }
	All shares $z_{jlt^0}$ are valid. 
\end{assumption}
\noindent Under the strict exclusion restriction and relevance, the shift-share IV estimator is consistent (Proposition 2 in GSS). Note that when Assumption \ref{ass:excls} is fulfilled, the shifts do not play a role for the validity of the instrument.



Another way to achieve consistency of the estimator is relying on random shifts. 
This is the setting in \citet*{Borusyak2020Quasi} and \citet*{Adao2019Shift}. 
Which of the settings should be considered is dependent on the application. 
Still, the methods proposed here are also applicable to the random shocks setting of \citet*{Borusyak2020Quasi}, when there are multiple shifts. The following sections discuss how this can be achieved.

Applications in labor and migration economics often are related to share exogeneity, because they stress that past shares are not directly related with the outcome of interest and are hence valid. Twenty-one examples for this are listed in Table \ref{tab:Quotes} in the Appendix. This list is not exhaustive. In the mentioned papers, the reader can find explicit statements that share exogeneity motivates the validity of the shift-share IV strategy.

\subsection{Violations of the exclusion restriction}\label{sec:violations}

The definition of validity in the preceding section makes it clear that violations of the exclusion restriction can come from two different sources. First, a non-zero $\alpha_j$ invalidates the shares. In the migration setting, \citet*{Jaeger2020Shift} warn that there might be direct effects through general equilibrium adjustments. The concern is that the economy reacts dynamically to migrant inflows. If this is the case, there is a direct correlation between instrument and outcomes, through native labor, capital and other general equilibrium adjustment channels, invalidating the instrument. One way that this might apply is illustrated by \citet*{Borjas2003Labor}: if migrants choose to move to regions with persistently high wages and native workers choose to migrate in response to the immigration of foreign workers, then the effect of immigration is positively biased.

Second, when unobserved shocks today and at the initial period $t^0$ are correlated ($\rho \neq 0$)  \textit{and} the initial shocks are related with initial shares ($\phi_j \neq 0$), this induces a non-zero correlation between instrument and error term. A violation is plausible, because serial correlation of unobservables is typically discussed in the literature \citep*[see Table \ref{tab:Quotes} and ][]{Jaeger2020Shift} and initial migrants might well have been attracted by economic conditions. 
In principle, the bias could go in both directions because migrants might endogenously select into regions with higher wages, or into regions with lower growth potential. 



The exclusion restriction is strict in the sense that it must hold for all $J$ shares. 
What looks like a single exclusion restriction in a just-identified model is in fact a set of $J$ exclusion restrictions. Therefore, the researcher needs to feel comfortable defending the exclusion restriction for Mexicans, Cubans, Canadians, Indians and all origin countries used when constructing the IV. 

In practice, it is very difficult, if not impossible to credibly uphold the strict exclusion restriction. 
While building an intuition about which shares are valid might be feasible, arguing that none of them had a long-term effect or was correlated with initial shocks is very restrictive. Thinking about which factors determined migrant settlement at an initial point in time makes it clear how difficult and hypothetical such an argument is destined to be. Institutional knowledge about which origin country group was mostly drawn into cities which were experiencing a boom at the time of settlement is typically unavailable. This holds true especially in settings in which a large number of countries of origin is used. Such detailed knowledge about the structural mechanisms at work is only available for very few countries, if any. 

Until now, there have not been attempts to make shift-share designs robust to violations of the exclusion restriction in Assumption \ref{ass:excls}. 
GSS propose computing sensitivity-to-misspecification (Rotemberg) weights, which indicate by what percentage the bias of the shift-share IV estimator changes if the bias from a certain share increases by one percent. The authors point out that one should argue prudently for the validity of shares associated with large weights. While these weights indicate the \textit{relative} importance with which an individual invalid share contributes to the bias of the estimator, the latter can still be considerable in \textit{absolute} terms, even if only shares associated with low weights violate the exclusion restriction. Therefore, it does not suffice to argue for the validity of the shares associated with the largest weights to make a case for a low bias in absolute terms.





\subsection{Some valid and some invalid share setting}

These reasons for violations of the strict exclusion restriction indicate that in many settings it can at best be hoped that \textit{some but not all} shares are valid. The general share validity setup as stated by GSS might be credible, but not for all shares.

\noindent The migration example applies to a setting with partial violation of the exclusion restriction for the following reasons. 
First, when $\rho\neq0$, some migrant groups might be related with labor demand shocks at the base year ($\phi_j\neq0$), while others are not. 
The absence of correlation with unobservable shocks is credible for some shares, because only some origin country groups might have migrated mainly because of economic reasons. This is in line with \citet{Jaeger2007Green}, who finds that migrants with employment visa where most responsive to economic conditions in their location choice. If the visa composition varies by origin groups, then some shares might have been driven mostly by factors orthogonal to economic conditions. \citet{Jaeger2007Green} also finds that in the beginning of the 1970s the share of employment-based visa was low. The increase of employment visa over the decades implies that origin country groups in a later base period are more likely to be invalid. 

Second, there are multiple sets of shares, which vary by base year. Some base years are correlated with the current shocks. Then for some years, $\rho_{0}\neq0$, while for others $\rho_{-1}=0$, when the correlation breaks after a few decades. 
Third, some origin country groups might have had long-term effects on wages, while the effects of others have worn off quickly. This might be the case when origin country shares which consisted mostly of people with family visas did not affect other factors of production in the long-term. 

In applications, the discussion of single shares as potentially problematic indicates that the researchers think of a setting in which some shares are valid, while others are invalid. Another telltale sign of such a setting is when researchers report Rotemberg weights and exclude the shares with the highest weights as a robustness exercise. The questions in the application of this diagnostic are: ``By how much does the bias of the estimator change, if a certain share is invalid? What happens if we assume that the most influential shares are invalid and exclude them from the estimation?'' These questions imply that it is feasible that some shares are valid, while others are invalid.

In the literature, there is also evidence that such a some valid - some invalid setting is indeed the case. 
\citet{Tabellini2020Gifts} raises the concern that specific origin country shares violate the exclusion restriction because Italian or Irish migrants could have chosen their city of location endogenously, based on the possibility to influence the local economy and politics. \citet*{Hunt2017Impact} and \citet*{Wozniak2012Timing} use adapted versions of the shift-share instrument where certain origin countries are excluded from the construction.

Arguably, the random shocks setting can be used when the strict exclusion restriction fails, but when the shocks are not numerous and random, which is often the case, this alternative approach is of little help. This offers a further setting where the some valid - some invalid IV setting applies. 
Several shift-share IVs can be constructed by using various push factors of emigration as shifts, such as economic, conflict- or civil liberties related variables. One might argue that economic variables are most likely to be related across countries, while political variables at origin are more likely to be unrelated with the local economic outcomes at destination. This setting is not based on the validity of shares and illustrates that the methods are in fact more widely applicable, also to the \citet*{Borusyak2020Quasi} context.

\section{Selection of valid IVs in shift-share estimation}\label{sec:modBartik}

In this section, I introduce how to obtain modified estimators which are robust to invalid shares. I present the general procedure, the leveraged methods and extensions of these methods. 

\subsection{Two-step procedure}

The idea of the procedure is to preselect valid shares beforehand with methods that will be presented in the following. I first introduce some notation. 
Let $\mathbf{Z}_\mathcal{V}$ be the matrix of valid IVs with $\mathcal{V} = \{j: \alpha_j = 0\}$ the set of valid IVs and $\hat{\mathcal{V}}$ the set of IVs \textit{selected} as valid.
Let $\mathbf{Z}_\mathcal{I}$ be the matrix of invalid IVs with $\mathcal{I} = \{j: \alpha_j \neq 0\}$ the set of invalid IVs and $\hat{\mathcal{I}}$ the set of IVs selected as invalid. 
Further, $|\mathcal{V}|$ is the number of valid and $|\mathcal{I}|$ is the number of invalid IVs. 

\noindent In short, the procedure works as follows:
\begin{enumerate}[topsep=0pt,itemsep=-1ex,partopsep=1ex,parsep=1ex]
	\item Use Adaptive Lasso or Confidence Interval Method with $\mathbf{y}$ as outcome, $\mathbf{x}$ as exposure and instrument with the share matrix $\mathbf{Z}$. The share matrix consists of all shares that the researcher believes to be valid.\footnote{The outcome and exposure are denoted by vectors $\mathbf{y}$ and $\mathbf{x}$, while the matrix $\mathbf{Z}$ collects the share instruments. Generally, scalars are in lower-case, vectors are in lower-case bold and matrices in upper-case bold.}
	\item Use shares chosen as valid (associated with $\alpha_j=0$) for constructing the corrected IV 
	\begin{equation}
	\sum_{j \in \hat{\mathcal{V}}} z_{ljt_0} \cdot g_{jt}
	\end{equation}
	and estimate the model with 
	\begin{enumerate}
		\item 2SLS or limited-information maximum likelihood with the selected shares
		\item the adjusted shift-share IV.
	\end{enumerate}
\end{enumerate}
It is important that the shares selected as invalid are controlled for. The invalid shares can only be omitted from the regression, if they are uncorrelated with the valid shares. However, this is unlikely to be the case in practice. Consistency of the proposed method follows directly if the selection methods used in the first step consistently select the invalid shares. 

When validity is plausible only with random shifts and there are multiple shifts, one can also apply an industry-level regression with multiple shifts and select shifts instead of shares, analogously to above. 
Disregarding which source of validity is put emphasis on, the preselection of variables starts with theoretical arguments. The set of shares (or shifts) selected is hence the intersection of the shares considered to be valid by the researcher and the algorithm. 

\subsection{Relaxed exclusion restriction}\label{sec:Assumptions}

In this section, I start with the critical assumptions needed for identification in the adaptive Least absolute shrinkage and selection operator (\textit{AL}) by \citet*{Windmeijer2019Use} and the Confidence Interval Method (\textit{CIM}) by \citet*{Windmeijer2020Confidence}, that I will use in this paper. Descriptions of the methods can be found in appendix \ref{app:MethodologicalAppendix} and in the original papers. 

The properties of the methods that will be leveraged to improve shift-share estimation are the so-called ``oracle properties''. Oracle properties mean consistent selection of invalid IVs and convergence in distribution to the ideal (\textit{oracle}) estimator that uses the model under perfect knowledge about the identity of invalid IVs. In Appendix \ref{app:Oracle} I describe the oracle estimator more closely. 

\begin{definition}
	\textbf{Oracle properties}
	\begin{enumerate}[topsep=0pt,itemsep=-1ex,partopsep=1ex,parsep=1ex]
		\item Consistent selection of invalid IVs: $\lim\limits_{n \rightarrow \infty} P(\hat{\mathcal{I}} = \mathcal{I}) = 1$
		\item Convergence in distribution: $\sqrt{n} (\hat{\beta} - \beta_0) 
		\overset{d}{\rightarrow} N(0, \sigma^2_{or})$,
	\end{enumerate}
	where $\sigma^2_{or}$ is the variance of the oracle estimator. 
\end{definition}
\noindent
In other words, if an estimator has oracle properties, it works as well as if one knew the true identity of invalid IVs. 
The AL has oracle properties when the majority of IVs is valid. All of the IVs also need to be relevant, as noted in equation \ref{eq:ModelFS-Separate}.
\begin{assumption}\label{ass:Majority}
	\textbf{Majority condition: } $
	|\mathcal{V}| > \frac{J}{2} \text{.}$
\end{assumption}
\noindent 
The Confidence Interval Method has oracle properties when the largest group of IVs is valid. 
The plurality condition in \citet*{Windmeijer2020Confidence} states that the group of valid IVs is larger than any other group. A group is defined as a set of IVs associated with an estimate which asymptotically deviates from the true $\beta$ by the same constant $c = \frac{\alpha_j}{\gamma_j}$. For the valid group, $c$ is zero. Formally, the plurality exclusion restriction is 
\begin{assumption}\label{ass:Plurality}\textbf{Plurality exclusion restriction: }
	$|\mathcal{V}| > max_{c \neq 0}| \{ j: \frac{\alpha_j}{\gamma_j} = c\} | \text{.}$
\end{assumption}
\noindent To compare these two assumptions, consider the following example: there are five IVs. The true effect is $\beta=1$. For three of these IVs: $\alpha_j=0$ and hence the three IV-specific estimands are $\beta_j = \beta$, while the remaining estimands are $\beta + \frac{\alpha_j}{\gamma_j}$ with $\alpha_j\neq0$. In this example: $\beta_1=\beta_2=\beta_3=1$ while e.g. $\beta_4=4$ and $\beta_5=5$. Clearly, the majority assumption is fulfilled. The plurality assumption is also fulfilled, because the largest group of IVs is valid. When only two IVs are valid and the third now has an estimand which is $\beta_3 = 3$, the majority is violated, because only 2/5 IVs are valid, but the plurality is still fulfilled, because there is one valid group of two IVs and three singleton groups. Therefore, the plurality assumption can still hold even when the majority is violated. 
Next, I discuss the choice of methods. I introduce how the two methods work in Appendices \ref{sec:AL} and \ref{sec:CIM}.

\subsection{Choice of methods}
The procedure builds on two methods from an emerging literature that investigates IV estimation in presence of invalid IVs. 
The proposed methods are the only ones which combine the following four benefits.

First, they are computationally feasible. \citet*{Andrews1999Consistent} requires to search over all possible models, which is computationally infeasible when the number of IVs is moderately large. 
Second, they do not require a priori knowledge about an initial set of valid IVs. \citet*{Caner2018Adaptive} also allow for invalid IVs when a set of valid IVs is known a priori. 
Third, the methods do not need assumptions on the correlation of first-stage and structural parameters.  
\citet*{Kolesar2015Identification} assume that first stage and direct effects are uncorrelated, but in applications, this assumption is rather strict. 
Finally, the direct effect of invalid IVs on the outcome need not be close to zero. This needs to be the case in \citet*{Conley2012Plausibly}, where additionally prior knowledge on possible values of $\alpha$ is needed. The methods used in this paper allow for arbitrarily strong direct effects. In fact, their performance even improves when the direct effects are large. 

In the following, I apply the methods to two real-world examples. 
I first reproduce the original estimates by using the standard shift-share IV 
which uses all shares, irrespectively of their validity. 
I then compare this regression with the result of the adjusted estimators, using AL and the confidence interval method. In the Appendix (Section \ref{sec:MCsim}) I also apply the methods in a Monte Carlo simulation that illustrates how the methods work with weaker IVs and strong violations.  

\section{Example 1: The effect of immigration on wages} \label{sec:BP}

\subsection{Setting}



The first empirical application is the estimation of the effect of immigration on wages in the United States. \citet*{Basso2015Association} estimate the linear model\footnote{
	I choose to use this paper as a reference even though it is unpublished, for the following reasons: the number of locations is large, which is helpful, because the methods I use make asymptotic arguments. Many of the published papers for which data is available have observation numbers which are low. For example, \citet{Card2009Immigration}, which GSS use as an illustration, uses only 124 city-observations.
}
\begin{equation}\label{eq:BP}
	\Delta y_{lt} = \beta \Delta immi_{lt} + \psi_{t} + \varepsilon_{lt} \text{,}
\end{equation}
with three time periods $t$ (1990, 2000, 2010) and 722 commuting zones $l$. On the left hand side, $\Delta y_{lt}$ stands for the three dependent variables used in separate regressions: the change in log weekly wages, and the change in log weekly wages of high- and low-skilled workers. On the right-hand-side, $\Delta immi_{lt}$ is the change in share of immigrants in total employment and $\beta$ is the coefficient of interest. Decade fixed-effects are denoted by $\psi_{t}$ and $\varepsilon_{lt}$ is the error term. Commuting zone fixed effects are accounted for by first-differencing.

As discussed before, estimating by OLS does not account for migrant sorting into regions: migrants might select into more prosperous or declining regions, creating a correlation between migrant location and outcome, which cannot be accounted to the impact of immigration. 
To tackle this problem, a shift-share IV, which uses origin-specific migrant shares in 1970 and changes in migrant populations, is used. The shift-share IV is  
\begin{equation}
	s_{lt} = \sum_c \left( z_{lc,1970} \cdot g_{ct} \right) \text{,}
\end{equation}
where $z_{lc,1970}$ is the share of immigrants from country $c$ in a location $l$ at base period 1970, and 19 origin country groups are used. The change of immigrants from country $c$ is denoted by $g_{ct}$. All of the origin countries are assumed to plausibly fulfill the exclusion restriction a priori. 

The SSIV estimates are in column 1 of \autoref{tab:BP}.  The coefficient estimate of the change in log weekly wages of the natives is 0.09, but it is insignificant. The 2SLS estimate is 0.479 ($p < 0.05$) and the LIML estimate is 0.568 ($p > 0.1$). For the change in log weekly wages of the high-skilled, the coefficients are 0.35 for SSIV, 0.519 for 2SLS and 0.672 for LIML. The coefficient is significant for 2SLS. For low-skilled workers, the effects on wages are negative at -0.66 and statistically significant for the shift-share analysis, and they are close to 0.1 for 2SLS and LIML. 
Overall, the standard estimates suggest  positive effects on high-skilled and null or negative effects on low-skilled wages. The first-stage F-statistics are at 171.3 for the overidentified models and at 21.8 for the SSIV. 

The possible violations in the migration context have been discussed in section \ref{sec:violations}. 
However, it is unclear whether these concerns really applied to some origin country groups and if yes to which. The remainder of this section shows the shares selected as invalid and how large the adjusted estimate is.

\afterpage{
	\newgeometry{left=0.6in, right=0.6in, top=1in, bottom=1in}
	\begin{table}[!htb]
		\begin{center}
			
			\caption{Impact of immigration on wages\label{tab:BP}}
			{
\def\sym#1{\ifmmode^{#1}\else\(^{#1}\)\fi}
\begin{tabular*}{\textwidth }{@{\hskip\tabcolsep\extracolsep\fill}l*{7}{C{2cm} C{2cm} C{2cm} C{2cm} C{2cm} C{2cm} C{2cm} C{2cm}}}
                    &\multicolumn{1}{c}{(1)}&\multicolumn{1}{c}{(2)}&\multicolumn{1}{c}{(3)}&\multicolumn{1}{c}{(4)}&\multicolumn{1}{c}{(5)}&\multicolumn{1}{c}{(6)}&\multicolumn{1}{c}{(7)}\\
                    &\multicolumn{1}{c}{Standard}&\multicolumn{1}{c}{AL (HS)}&\multicolumn{1}{c}{CIM (HS)}&\multicolumn{1}{c}{AL (HS)}&\multicolumn{1}{c}{CIM (HS)}&\multicolumn{1}{c}{AL (AR)}&\multicolumn{1}{c}{CIM (AR)}\\
\toprule \multicolumn{8}{c}{\textit{Panel A}: Change in average log weekly wages} \\ \midrule SSIV&      0.0921         &      0.0921         &      0.0921         &      0.0921         &      0.0921         &       0.625\sym{+}  &      0.0176         \\
                    &     (0.438)         &     (0.438)         &     (0.438)         &     (0.438)         &     (0.438)         &     (0.336)         &     (0.438)         \\
F                   &       21.80         &       21.80         &       21.80         &       21.80         &       21.80         &       11.30         &       16.08         \\
\end{tabular*}
}

			{
\def\sym#1{\ifmmode^{#1}\else\(^{#1}\)\fi}
\begin{tabular*}{\textwidth }{@{\hskip\tabcolsep\extracolsep\fill}l*{7}{C{2cm} C{2cm} C{2cm} C{2cm} C{2cm} C{2cm} C{2cm} C{2cm}}}
\midrule
2SLS                &       0.479\sym{*}  &       0.479\sym{*}  &       0.479\sym{*}  &       0.479\sym{*}  &       0.479\sym{*}  &       0.189\sym{*}  &      -0.287         \\
                    &     (0.195)         &     (0.195)         &     (0.195)         &     (0.195)         &     (0.195)         &    (0.0798)         &     (0.206)         \\
F                   &       171.3         &       171.3         &       171.3         &       171.3         &       171.3         &       148.3         &       12.08         \\
\end{tabular*}
}

			{
\def\sym#1{\ifmmode^{#1}\else\(^{#1}\)\fi}
\begin{tabular*}{\textwidth }{@{\hskip\tabcolsep\extracolsep\fill}l*{7}{C{2cm} C{2cm} C{2cm} C{2cm} C{2cm} C{2cm} C{2cm} C{2cm}}}
\midrule
LIML                &       0.568         &       0.568         &       0.568         &       0.568         &       0.568         &       0.183\sym{*}  &      -0.367         \\
                    &     (0.445)         &     (0.445)         &     (0.445)         &     (0.445)         &     (0.445)         &    (0.0809)         &     (0.237)         \\
F                   &       171.3         &       171.3         &       171.3         &       171.3         &       171.3         &       148.3         &       12.08         \\
\midrule \# inv     &           0         &           0         &           0         &           0         &           0         &           9         &          10         \\
Sign.               &           -         &        0.05         &        0.05         &         0.1         &         0.1         &     0.01302         &     0.01302         \\
\end{tabular*}
}
		
			{
\def\sym#1{\ifmmode^{#1}\else\(^{#1}\)\fi}
\begin{tabular*}{\textwidth }{@{\hskip\tabcolsep\extracolsep\fill}l*{7}{C{2cm} C{2cm} C{2cm} C{2cm} C{2cm} C{2cm} C{2cm} C{2cm}}}
\midrule \multicolumn{8}{c}{\textit{Panel B}: Change in avg. log weekly wages of high-skilled} \\ \midrule SSIV&       0.347         &       0.347         &       0.347         &       0.347         &       0.347         &       1.062\sym{*}  &       0.268         \\
                    &     (0.299)         &     (0.299)         &     (0.299)         &     (0.299)         &     (0.299)         &     (0.447)         &     (0.294)         \\
F                   &       21.80         &       21.80         &       21.80         &       21.80         &       21.80         &       26.17         &       15.15         \\
\end{tabular*}
}

			{
\def\sym#1{\ifmmode^{#1}\else\(^{#1}\)\fi}
\begin{tabular*}{\textwidth }{@{\hskip\tabcolsep\extracolsep\fill}l*{7}{C{2cm} C{2cm} C{2cm} C{2cm} C{2cm} C{2cm} C{2cm} C{2cm}}}
\midrule
2SLS                &       0.519\sym{**} &       0.519\sym{**} &       0.519\sym{**} &       0.519\sym{**} &       0.519\sym{**} &       0.180\sym{+}  &      -0.529\sym{**} \\
                    &     (0.173)         &     (0.173)         &     (0.173)         &     (0.173)         &     (0.173)         &    (0.0973)         &     (0.171)         \\
F                   &       171.3         &       171.3         &       171.3         &       171.3         &       171.3         &       6.557         &       22.28         \\
\end{tabular*}
}

			{
\def\sym#1{\ifmmode^{#1}\else\(^{#1}\)\fi}
\begin{tabular*}{\textwidth }{@{\hskip\tabcolsep\extracolsep\fill}l*{7}{C{2cm} C{2cm} C{2cm} C{2cm} C{2cm} C{2cm} C{2cm} C{2cm}}}
\midrule
LIML                &       0.672         &       0.672         &       0.672         &       0.672         &       0.672         &       0.164         &      -0.604\sym{**} \\
                    &     (0.503)         &     (0.503)         &     (0.503)         &     (0.503)         &     (0.503)         &     (0.103)         &     (0.188)         \\
F                   &       171.3         &       171.3         &       171.3         &       171.3         &       171.3         &       6.557         &       22.28         \\
\midrule \# inv     &           0         &           0         &           0         &           0         &           0         &          11         &           9         \\
Sign.               &           -         &        0.05         &        0.05         &         0.1         &         0.1         &     0.01302         &     0.01302         \\
\end{tabular*}
}
		
			{
\def\sym#1{\ifmmode^{#1}\else\(^{#1}\)\fi}
\begin{tabular*}{\textwidth }{@{\hskip\tabcolsep\extracolsep\fill}l*{7}{C{2cm} C{2cm} C{2cm} C{2cm} C{2cm} C{2cm} C{2cm} C{2cm}}}
\midrule \multicolumn{8}{c}{\textit{Panel C}: Change in avg. log weekly wages of low-skilled} \\ \midrule SSIV&      -0.655\sym{+}  &      -0.655\sym{+}  &      -0.641\sym{+}  &      -0.655\sym{+}  &      -0.657\sym{+}  &      -0.721\sym{+}  &      -0.151         \\
                    &     (0.345)         &     (0.345)         &     (0.344)         &     (0.345)         &     (0.368)         &     (0.378)         &     (0.205)         \\
F                   &       21.80         &       21.80         &       19.92         &       21.80         &       16.89         &       13.90         &       9.101         \\
\end{tabular*}
}

			{
\def\sym#1{\ifmmode^{#1}\else\(^{#1}\)\fi}
\begin{tabular*}{\textwidth }{@{\hskip\tabcolsep\extracolsep\fill}l*{7}{C{2cm} C{2cm} C{2cm} C{2cm} C{2cm} C{2cm} C{2cm} C{2cm}}}
\midrule
2SLS                &       0.129         &       0.129         &       0.185         &       0.129         &       0.162         &     -0.0684         &      -0.501         \\
                    &     (0.160)         &     (0.160)         &     (0.189)         &     (0.160)         &     (0.188)         &    (0.0793)         &     (0.699)         \\
F                   &       171.3         &       171.3         &       156.9         &       171.3         &       138.6         &       88.74         &       0.650         \\
\end{tabular*}
}

			{
\def\sym#1{\ifmmode^{#1}\else\(^{#1}\)\fi}
\begin{tabular*}{\textwidth }{@{\hskip\tabcolsep\extracolsep\fill}l*{7}{C{2cm} C{2cm} C{2cm} C{2cm} C{2cm} C{2cm} C{2cm} C{2cm}}}
\midrule
LIML                &      0.0745         &      0.0745         &       0.135         &      0.0745         &       0.119         &     -0.0825         &      -4.118         \\
                    &     (0.226)         &     (0.226)         &     (0.324)         &     (0.226)         &     (0.274)         &    (0.0821)         &     (18.82)         \\
F                   &       171.3         &       171.3         &       156.9         &       171.3         &       138.6         &       88.74         &       0.650         \\
\midrule \# inv     &           0         &           0         &           1         &           0         &           4         &           9         &          15         \\
Sign.               &           -         &        0.05         &        0.05         &         0.1         &         0.1         &     0.01302         &     0.01302         \\
\bottomrule          \end{tabular*} }

			\floatfoot{\footnotesize \textit{Note:} This table reports estimates of $\beta$ in equation \ref{eq:BP}. $N= 2166$ (722 CZ $\times$ 3). Standard errors (in parentheses) are clustered by commuting zone. First-stage F-statistics are reported. Observations are weighted by beginning-of-period population. Outcome variables are listed in the panel heads. In the first row of each panel, shift-share IV results are reported. In the second and third rows, the full vector of shares is used for 2SLS and LIML. In the last rows, the number of countries chosen as invalid and the thresholds used in the HS procedure are reported. In column (1), all shares are assumed to be valid. Column heads of columns 2 to 5 denote which method has been used for selection.}
		\end{center}
	\end{table}
	\clearpage
	\restoregeometry
}

\afterpage{
	\begin{landscape}
		\begin{table}[!htb]
			\caption{Countries selected as invalid}\label{tab:ExclCountriesBP}
			\begin{tiny}
	\begin{tabular}{*{22}{c}}
		DV & Method & SL & \rot{CanAL} & \rot{Rest of Americas} &\rot{Mexico} &\rot{Scandinavia and Northern Europe} &\rot{UK and Ireland} &\rot{China} &\rot{Japan} &\rot{Korea} &\rot{Philippines} &\rot{Vietnam} &\rot{India} &\rot{Western Europe} &\rot{Africa} &\rot{Oceania} &\rot{Other} &\rot{Mediterranean Countries} &\rot{Central and Eastern Europe} &\rot{Baltic States} &\rot{Rest of Asia}\\
		\toprule
		\multicolumn{22}{c}{Single regressor}\\
		\midrule
		dlweekly & AL & 0.01302 & - & - & - & - & - & - & - & - & - & - & - & - & - & - & - & - & - & - & - \\
		dlweekly & CIM & 0.01302 & - & - & - & - & - & - & - & - & - & - & - & - & - & - & - & - & - & - & - \\
		dlweekly & AL & 0.05 & - & - & - & - & - & - & - & - & - & - & - & - & - & - & - & - & - & - & - \\ 
		dlweekly & CIM & 0.05 & - & - & - & - & - & - & - & - & - & - & - & - & - & - & - & - & - & - & - \\
		dlweekly & AL & 0.1 & - & - & - & - & - & - & - & - & - & - & - & - & - & - & - & - & - & - & - \\
		dlweekly & CIM & 0.1 & - & - & - & - & - & - & - & - & - & - & - & - & - & - & - & - & - & - & - \\
		dlweekly & AL AR & 0.01302 & x & - & x & - & x & x & - & x & x & - & - & - & - & - & - & x & x & - & x\\
		dlweekly & CIM AR & 0.01302 & - & - & - & - & - & - & x & - & x & - & x & x & x & x & x & x & x & - & x \\
		dlweekly\_hskill & AL & 0.01302 & - & - & - & - & - & - & - & - & - & - & - & - & - & - & - & - & - & - & - \\
		dlweekly\_hskill & CIM & 0.01302 & - & - & - & - & - & - & - & - & - & - & - & - & - & - & - & - & - & - & - \\
		dlweekly\_hskill & AL & 0.05 & - & - & - & - & - & - & - & - & - & - & - & - & - & - & - & - & - & - & - \\
		dlweekly\_hskill & CIM & 0.05 & - & - & - & - & - & - & - & - & - & - & - & - & - & - & - & - & - & - & - \\
		dlweekly\_hskill & AL & 0.1 & - & - & - & - & - & - & - & - & - & - & - & - & - & - & - & - & - & - & - \\
		dlweekly\_hskill & CIM & 0.1 & - & - & - & - & - & - & - & - & - & - & - & - & - & - & - & - & - & - & - \\
		dlweekly\_hskill & AL AR & 0.01302 & - & x & x & - & x & x & - & x & x & - & - & - & x & x & - & x & x & - & x\\
		dlweekly\_hskill & CIM AR & 0.01302 & - & - & - & - & - & - & x & - & x & - & x & x & x & x & x & - & x & - & x\\
		dlweekly\_lskill & AL & 0.01302 & - & - & - & - & - & - & - & - & - & - & - & - & - & - & - & - & - & - & - \\
		dlweekly\_lskill & CIM & 0.01302 & - & - & - & - & - & - & - & - & - & - & - & - & - & - & - & - & - & - & - \\
		dlweekly\_lskill & AL & 0.05 & - & - & - & - & - & - & - & - & - & - & - & - & - & - & - & - & - & - & - \\
		dlweekly\_lskill & CIM & 0.05 & - & - & - & - & - & - & - & - & - & - & x & - & - & - & - & - & - & - & - \\
		dlweekly\_lskill & AL & 0.1 & - & - & - & - & - & - & - & - & - & - & - & - & - & - & - & - & - & - & - \\
		dlweekly\_lskill & CIM & 0.1 & - & - & - & - & - & - & - & - & x & - & - & - & x & - & - & - & x & - & x \\
		dlweekly\_lskill & AL AR & 0.01302 & x & - & - & - & x & x & x & x & x & - & - & - & x & - & - & - & x & - &  x \\
		dlweekly\_lskill & CIM AR & 0.01302 & x & - & x & x & - & x & x & x & x & - & x & x & x & - & x & x & x & x & x \\
		\midrule
		\multicolumn{22}{c}{Multiple regressors}\\
		\midrule
		dlweekly & AL HS & 0.1 & -&-&-&-&-&-&-&-&-&-&-&-&-&-&-&-&-&-&- \\
		dlweekly & AL AR & 1970 & - & - & - & - & - & - & - & - & - & - & - & - & - & - & - & - & x & - & x \\
		&  & 1980 & x & x & - & x & x & - & - & - & - & x & - & - & x & - & x & - & - & - & x \\
		dlweekly\_hskill & AL HS & 1970 & - & - & - & - & x & - & - & - & - & - & - & - & - & - & - & - & - & - & -\\
		&  & 1980 & - & - & - & - & x & - & - & - & - & - & - & - & - & - & - & - & - & - & -\\
		dlweekly\_hskill & AL AR & 1970 & - & - & - & - & x & - & - & - & - & - & - & - & - & - & - & - & x & - & x\\
		&  & 1980 & x & x & x & x & x & - & - & - & - & - & x & - & x & - & x & - & - & - & -\\
		dlweekly\_lskill & AL HS & 0.1 & -&-&-&-&-&-&-&-&-&-&-&-&-&-&-&-&-&-&- \\
		dlweekly\_lskill & AL AR & 1970 & - & x & - & - & - & - & - & - & - & - & - & - & - & - & - & - & x & - & x \\
		&  & 1980 & x & x & - & x & x & x & - & - & - & - & - & - & - & - & - & x & - & - & - \\
		\bottomrule
		\multicolumn{22}{c}{%
			\begin{minipage}{\textwidth- 35 \tabcolsep}%
				\vspace{2px}
				\textit{Note:} This table reports the countries chosen as invalid in tables \ref{tab:BP-SSIV} and \ref{tab:BP-mult-SSIV} for the reanalysis of \citet*{Basso2015Association}. The left columns display the method and the outcome variable used. x denotes a country selected as invalid.
			\end{minipage}%
		}
	\end{tabular}
\end{tiny}
		\end{table}
	\end{landscape}
}

\subsection{Results}
The results of applying AL and CIM on the immigration example are in Table \ref{tab:BP}. Each panel presents the results for one of the outcomes. A list of selected countries can be found in table \ref{tab:ExclCountriesBP}. 
Overall, with the new methods, the coefficients of immigration decrease and often switch sign. The decrease in coefficients tends to be stronger when CIM is used in the selection step.

When choosing the significance level of $0.1/ln(N)$ (0.01375) in the downward testing procedure as proposed in WLHB, no country is chosen as invalid. Thus, the adjusted estimators are identical to the original ones (col. 1). \citet*{Bowsher2002testing} shows that the use of many IVs leads to low power when using the HS test. Larger significance levels of the HS test are more conservative, which is the inverse logic as with conventional tests of coefficient significance \citep*{Roodman2009note}.  Hence, a more conservative strategy would be to set the threshold to a more conventional level, for example to $0.05$ or $0.1$. Increasing this threshold in the testing procedure leads to the selection of a few countries for low-skilled wages (Panel C), but does not change the results qualitatively. 

One might be concerned that the IVs are weak and hence the HS test is unreliable. To address this concern, I also use the Anderson-Rubin test in the downward selection procedure. As a threshold I use $0.1/ln(N)$, as originally proposed in WFDS. 
Now, both methods select many IVs. 
Often, more than half of shares are selected as invalid. If a majority is invalid, AL in fact does not have oracle properties. I therefore rely on the CIM. The preferred analyses are hence those in the last column of table \ref{tab:BP} for 2SLS and LIML. All estimates decrease strongly, and mostly become negative.


For overall weekly wages, the coefficients become negative and are statistically insignificant. 
For wages of the high-skilled, the estimates from overidentified models become negative and statistically significant when selecting via CIM, which is in stark contrast with the original results. Interestingly, the absolute size of coefficients is very similar to the original ones, with the difference that they changed direction. 
For wages of the low-skilled, now 15 countries are chosen as invalid by CIM. The coefficients of 2SLS and LIML are negative but none of them are significant. This might be because the F-statistic becomes low. Still, for most other analyses the F-statistics are still reasonably high.

It is reassuring to see that the differences between 2SLS and LIML estimators are smaller with the corrected as compared to the standard estimators. 
The remaining differences in the adjusted shift-share (SSIV), 2SLS and LIML estimates may stem from different reasons. First, LIML approximately eliminates the finite sample bias that is due to weak instruments when using 2SLS.  
Second, the weighting scheme of each just-identified IV estimate differs across SSIV and 2SLS. The weights shown in GSS are dependent on the shift variable. The weighting implicit in the 2SLS estimator which uses only shares, does not take shifts into account. Therefore, different results may also arise because different methods estimate different weighted combinations of just-identified estimates.





\subsection{Results for dynamic effects}

Taking into account the critique of \citet*{Jaeger2020Shift}, who argue that using a single regressor compresses the long- and short-term effects, means to include lagged migration. The equation now becomes
\begin{equation}\label{eq:BP-dynamic}
	\Delta y_{lt} = \beta_c \Delta immi_{lt} + \beta_l \Delta immi_{lt-1} + \psi_{t} + \varepsilon_{lt} \text{,}
\end{equation}
where $\beta_c$ is the coefficient of interest for the contemporaneous impact and $\beta_l$ denotes the coefficient of interest for the lagged impact. I include an additional shift-share IV, now using 1980 as a base period. When using 2SLS or LIML, the number of shares increases to 38 (19 per base year).

\noindent The results are shown in Table \ref{tab:BP-mult}. 
The standard estimates always suggest positive effects in the short and negative effects in the long run, across estimators and outcomes. This is exactly the opposite of what \citet*{Jaeger2020Shift} expect: partial equilibrium effects should be negative and general equilibrium adjustments are expected to offset these negative effects. 
These unexpected coefficient estimates might be due to the same endogeneity problems as before: both base years of the number of foreign borns might be directly correlated with the endogenous treatment. 

Using the extension of the adaptive Lasso presented in Appendices \ref{sec:AL} and \ref{sec:Extensions}, with the Hansen-Sargan downward testing procedure, only for wages of the high-skilled the UK and Ireland are selected as invalid once and the adjusted estimates still have the same signs.

When using the Anderson-Rubin test instead, nine to eleven countries are selected for each outcome. The countries selected have a large overlap. Most variables selected come from the year 1980. I focus on 2SLS and LIML results, because the Cragg-Donald statistic of the SSIV is very low. The estimates now have the expected sign: the coefficient of contemporaneous immigration is negative and that of lagged immigration is positive.

\subsection{Overidentification from multiple shifts}\label{sec:MultipleShiftsMigration}

One might fundamentally question the share exogeneity interpretation of the shift-share design in the migration setting. In principle all shares could directly affect wages. 
If this is the case, the shift-share IV can be motivated via random shifts as in \citet*{Borusyak2020Quasi}. This offers an alternative starting point for the selection methods. This shows that the new methods are not restricted to the exogenous share world of \citet*{Goldsmith-Pinkham2020Bartik}.

If validity was still a concern for all shares, one additional way to check for robustness of the results is to motivate the exclusion restriction through quasi-random shifts and to use country-of-origin specific push factors related to war, civil liberties or natural disasters. One example for such an approach is \citet*{Llull2017effect}. The selection methods can then be used with the different shifts in an overidentified model. There would be reason to believe that some instruments are valid while others are invalid. Some shifts are related to war, others to politics, again others to other country-of-origin factors. Some might be correlated with unobservable shocks which drive wages at destination, for others it is difficult to think of a reason why that would be the case. 

If multiple shifts are available, multiple shift-share instruments can be generated and used in an over-identified model. The SSIV constructed with shocks that fulfill the conditions in \citet*{Borusyak2020Quasi} can then be selected. Alternatively, one could also directly use the class-level regression, and use the shocks as IVs directly. The latter approach will be used in the international trade application. \citet*{Borusyak2020Quasi} show consistency of the IV-estimator, taking into account that the data is non-iid. This does not pose a challenge for the selection methods, because the key assumption is that a large-enough group of IV-specific estimators is consistent, regardless of how that consistency is established. 

Moreover, selection of a particular group does not necessarily mean that the other groups are invalid instruments. Different shocks might produce heterogeneous effects. The migration inflow due to war might be different from that due to a decrease in civil liberties, which is more likely to induce migration of the elite. 

To illustrate this approach, I used eleven shifts and produced eleven shift-share instruments, still using the 1970 country shares.\footnote{The shifts and their sources are as follows: migration (as in the preceding subsections), battle-related deaths, onesided violence and nonstate violence (Uppsala Conflict Data program, www.ucdp.uu.se), population (World Development Indicators), Civil Liberties, Political Rights, Freedom House Status \citep{House2020Freedom}, Polity Score (Polity V project), Press Freedom Status and Press Freedom Score \citep{House2017Freedom}.} I directly estimate the dynamic model suggested by \citet{Jaeger2020Shift}, including lagged immigration. 
My findings can be found in table \ref{tab:BP-Shifts2}. In brief, the main results stay the same: with the HS-testing procedure, only few IVs are selected as invalid, while with the Anderson-Rubin test, more IVs are selected. With the AR testing procedure, all estimates turn negative but insignificant. This could be due to a loss in relevance, as the first-stage F-statistic becomes low. The shifts selected as invalid can be found in table \ref{tab:ExclShocksBP}. The variables that are selected most often are the IVs constructed with battle-related deaths, with the political freedom indicator (every analysis) and with the Press Freedom Score (five times). Since the last two express similar things, it makes sense that they constitute a group.  

\subsection{Discussion}
There are five key takeaways from the application of AL and CIM to the estimation of the effect of immigration on wages. 
First, the results from adjusted estimators suggest a strong positive bias of standard estimates. This is in line with most of the literature, that expects an upward bias. This doesn't seem to be due to weaker instruments after selection, because the first-stage statistics are still reasonably high and the use of the LIML estimator, which has better finite-sample properties in presence of weak IVs suggests the same direction of the bias.


Second, the selection of shares is consistent with economic intuition. 
The selection of Central and Eastern Europe (including Russia) in almost all analyses can be explained by the emigration from the Soviet Union in the 1970s and the Post-Soviet countries in the 1990s. The emigrants predominantly chose coastal cities which had large country-of-origin communities, but also cities which had experienced lasting prosperity. The conditions which have made these places attractive might be correlated over time. This makes a violation of the exclusion restriction likely. 
The share of migrants from the UK and Ireland, which has been picked by \citet{Tabellini2020Gifts} as an example for possibly invalid shares is chosen nine times. 
When shares from multiple base years are used, mostly IVs with base year 1980 are selected.
This is consistent with more job-related visa in 1980 as compared to more family-related visa 1970 and is in line with the common practice in the literature of choosing longer lags to break eventual correlation between shares and current unobservable shocks.


Third, the application shows the added value of the methods to existing econometric tools. The Rotemberg weights proposed by GSS help understand which share's invalidity is most likely to bias results, but it does not tell the researcher whether this bias is large in absolute terms and it does not lend guidance on which country should be excluded effectively. 
A few of the countries flagged as potentially problematic by high weights have been selected. The Philippines have received the highest sensitivity-to-misspecification weight in GSS. Indeed, they have been selected seven times by AL and CIM, and adjusting for them results in large qualitative changes of the coefficients. 

However, if the Rotemberg weights for some origin countries are low, their invalidity could still contribute to a large part of the inconsistency of estimators. Notably, many country groups which are not worrisome according to the top-5 Rotemberg weights, such as Central and Eastern Europe have been chosen as invalid, while some that have high weights have not been selected. This shows how the new methods can guide the selection of shares beyond the discretion of researchers.

Fourth, the fraction of shares selected as invalid can be high. A maximum of 15 out of 19, are selected as invalid suggesting that the majority assumption is likely to be violated. This suggests that the adaptive Lasso can not consistently select valid IVs in the migration setting with one regressor. Also, there is a large overlap of selected countries as invalid, by variables and methods used. This is reassuring in that it confirms that the share selection is not erratic. 

Fifth, when including lagged immigration, the coefficient estimates have the expected sign, only with the proposed AL adjustment. The origin-country variables selected are mostly those from the year 1980. This is consistent with \citet*{Jaeger2020Shift}, who worry that spatial adjustments might take around ten years\footnote{``Research on regional evolutions in the U.S. concludes, however, that spatial adjustments can take around a decade or more.'' (p. 10)}. In my analysis, I also use data from 1990. Hence, my analysis confirms \cites{Jaeger2020Shift} result that using contemporaneous and lagged immigration can help uncover the effects of immigration. It also confirms the common practice of taking longer lags of the country-of-origin distribution to plausibly fulfill the exclusion restriction.

\section{Example 2: The China Shock}\label{sec:ADH}


\subsection{Setting}

\citet*[][ADH]{Autor2013china} study the impact of Chinese imports on employment in manufacturing in the US. 
The regression equation is 
\begin{equation}\label{eq:ADH}
	\Delta L_{lt}^m = \Delta IPW_{ult} \beta_1  + X_{lt}'\beta + \gamma_t + \varepsilon_{lt} \text{,}
\end{equation}
where the left-hand side is decadal change in manufacturing employment in commuting zone $l$, $\beta_1$ is the coefficient of interest and $\Delta IPW_{ult}$ is import exposure, defined as $\Sigma_{j} z_{ljt} g_{jt}$. Here, $z_{ljt}$ are the shares of workers in commuting zone $l$ employed in industry $j$ at time $t$ and $g_{jt}$ measures the growth of imports from China in industry $j$. This regression is estimated in first-differences to exclude commuting-zone fixed effects and augmented by a time dummy and a set of commuting-zone-level controls. The time period used ranges from 1990 to 2007 and there are  397 industry shares, indexed by four-digit SIC codes. 

The endogeneity issue that affects this analysis is that both employment and imports might be correlated with unobserved shocks to US demand. 
To address this problem, a shift-share instrument is used, which replaces the share of workers with the same share ten years earlier and uses import exposure of other high-income countries rather than the US. 
ADH find a coefficient of -0.596. I report the same coefficient for the original estimate in row 1, column 1 (1,1) of table \ref{tab:ADH}. When using all shares separately in a 2SLS estimation, a lower coefficient of -0.183 is found (2,1). The same model is also estimated by LIML (3,1). These are the baseline coefficients to which the adjusted estimation results will be compared. 

\subsection{Results}
One might understand the analysis from the viewpoint of  GSS in the framework of a pooled exposure research design, in which employment shares capture local exposure to common import shocks. 
My results show which industry shares one should worry about if one chooses to rely on share validity. 

ADH discuss the possible invalidity of three specific industries: the computer industry, construction materials as well as apparel, footwear and textiles.  GSS show that electronic computers display the highest sensitivity-to-misspecification weight, making the validity of this specific share especially important. 

\afterpage{
	\begin{landscape}
		\vspace{13cm}
		\begin{center}
			\begin{table}[htb]
				\caption{Impact of Chinese Import Exposure\label{tab:ADH}}
				\begin{tabular}{l*{7}{p{2cm} p{2cm} p{2cm} p{2cm} p{2cm} p{2cm} p{2cm} p{2cm}}}
                    &\multicolumn{1}{c}{(1)}&\multicolumn{1}{c}{(2)}&\multicolumn{1}{c}{(3)}&\multicolumn{1}{c}{(4)}&\multicolumn{1}{c}{(5)}&\multicolumn{1}{c}{(6)}&\multicolumn{1}{c}{(7)}\\
                    &\multicolumn{1}{c}{Original}&\multicolumn{1}{c}{ALasso}&\multicolumn{1}{c}{ALasso}&\multicolumn{1}{c}{CIM}&\multicolumn{1}{c}{CIM}&\multicolumn{1}{c}{AL AR}&\multicolumn{1}{c}{CIM AR}\\
\toprule
SSIV                &      -0.596&      -0.596&      -0.603&      -0.596&      -0.692&      -0.720&      -0.924\\
                    &    (0.0988)&    (0.0988)&    (0.0990)&    (0.0988)&     (0.125)&    (0.0781)&     (0.174)\\
F                   &       47.64&       47.64&       47.52&       47.64&       37.64&       71.83&       28.68\\
\end{tabular}

				\begin{tabular}{l*{7}{p{2cm} p{2cm} p{2cm} p{2cm} p{2cm} p{2cm} p{2cm} p{2cm}}}
\midrule
2SLS                &      -0.183&      -0.183&      -0.201&      -0.183&      -0.173&      -0.440&       0.117\\
                    &    (0.0419)&    (0.0419)&    (0.0420)&    (0.0419)&    (0.0485)&    (0.0527)&    (0.0813)\\
F                   &       69.86&       69.86&       69.86&       69.86&       37.75&       75.71&       44.82\\
\end{tabular}

				\begin{tabular}{l*{7}{p{2cm} p{2cm} p{2cm} p{2cm} p{2cm} p{2cm} p{2cm} p{2cm}}}
\hline LIML         &      -2.742&      -2.742&      -1.032&      -2.742&      -7.503&      -0.925&       2.017\\
                    &     (75.48)&     (75.48)&     (4.883)&     (75.48)&    (1119.3)&     (0.355)&     (3.852)\\
F                   &       69.86&       69.86&       69.86&       69.86&       37.75&       75.71&       44.82\\
\midrule \# inv     &           0&           0&           1&           0&           8&          63&         128\\
Sign.               &            &     0.01375&        0.05&     0.01375&        0.05&     0.01375&     0.01375\\
\bottomrule         \end{tabular}

				\floatfoot{\textit{Note:} This figure reports the estimates of $\beta_1$ in equation \ref{eq:ADH}. $N= 1444$ (722 CZ $\times$ 2). Column 1: Results when all shares assumed as valid, Columns 2 - 4: Endogenous shares selected by AL, Column 5: AL selection by SIC2-class, Columns 6 - 8: IV selection with Confidence Interval Method. First row of results: just-identified model with shift-share IV, Second row: 2SLS using all shares separately, Third row: LIML, using shares separately. Standard errors (in parentheses) are clustered by state and observations weighted by start of period CZ share of national population.}
			\end{table}
		\end{center}
	\end{landscape}
}

The results of the AL-adjusted IV estimators are presented in table \ref{tab:ADH}. 
Using AL, the coefficients change by little. With the default threshold of the over-identification test at $0.01375$ ($0.1/ln(N)$) as in WFDS, the test does not reject the Null hypothesis, all shares can be used for the construction of the shift-share IV and all coefficients are identical to the original estimates (column 2 of table \ref{tab:ADH}). 
To account for the problem of too many instruments in the HS-test, I set the threshold to $0.05$. 
Now, only one industry is selected. When excluding this industry from the construction of the instrument in column 3, the estimate is virtually unaltered. 

When applying CIM, the industry chosen by AL and seven additional industries are selected as invalid. The estimates becomes larger in absolute terms but the confidence interval still includes the original estimate. Hence, the application is also robust to omitting shares chosen as invalid. 

When estimating the post-selection model by LIML, the estimates are very different from 2SLS. This might indicate that many IVs are weak and 2SLS is therefore biased. In order to adjust for this, I use the Anderson-Rubin test in the downward testing procedure instead of the HS-test. 
When using adaptive Lasso with the AR-test, the method now selects 63 shares as invalid. When selecting with CIM and the AR-test (column 7 of table \ref{tab:ADH}), even 128 shares are chosen as invalid.  
The estimate for SSIV now moves to -0.92, but the 95\% significance interval still includes the original estimate. The 2SLS estimate becomes positive, with a coefficient of 0.12, while the coefficient of LIML is positive and large. 

The industries chosen as invalid are listed in table \ref{tab:ExclCountries} of the supplementary material. 
These industries concord with those discussed in ADH. The first industry labeled as problematic was the computer industry. Industries belonging to electronic and computer equipment (SIC35 and 36) are among those chosen most often, constituting up to 29 percent of the shares selected as invalid. The second industry class that is discussed in ADH is related to construction. Up to 16 percent of selected shares come from industries that are associated with construction (32, 33, 34). The third industry discussed in ADH is apparel, footwear and textiles. Also, up to 16 percent of shares selected comes from these industries (22, 23, 31). 

The analysis offers additional information beyond the sensitivity to misspecification illustrated by Rotemberg weights. 
Games, Toys and Children Vehicles as well as Household Audio and Video Equipment have obtained the second- and third-largest Rotemberg weights in GSS, and they have been selected by AL and CIM. The industry with fourth-largest Rotemberg has been selected by CIM and the one with the fifth-largest weight has been selected by both methods. However, the SIC-4 industry with the largest weight has not been selected, while numerous industries from the SIC-2 industry related to it and SIC-codes from the food sector have been selected. This shows how the proposed methods can guide share selection in expected ways but it can also inspire to think about the possible endogeneity of some other industries. 

Overall, if one believes that some shares are valid and some invalid the selection procedures single out industries which are also in harmony with the ones discussed by ADH. The results are relatively robust to the use of the new methods. 

\subsection{Overidentification from multiple shifts}

If share exogeneity is not credible, one can also understand exogeneity of the shift-share instrument from a random shift perspective, as in \citet*{Borusyak2020Quasi}, who run an equivalent industry-level regression which uses the shift-variable as instrument. In table C4 of their paper, they use an overidentified model with all eight shifts from other high-income countries instead of the aggregated shift.\footnote{In this analysis, the authors add lagged sum of shares for each period as control variables. I follow this modeling choice to keep results comparable.} 
The common concern is usually that imports of high-income countries are correlated with unobservable shocks. The estimates for these estimations lie at roughly -0.24 for both 2SLS and LIML. It is reassuring to see that the coefficients of the two methods coincide.

I reproduce the results from this table in unreported estimations. The HS- and AR-tests do not reject at any conventional significance level, and therefore the selection algorithms select all eight shifts as valid. 
This robustness to the use of the new methods illustrates how in this example identification should be thought of in terms of shifts. This is in line with \cites{Borusyak2020Quasi} and \cites{Adao2019Shift} interpretation of the exclusion restriction as shock exogeneity.

Researchers can also leverage a larger set of import shifts from even more countries in which they are confident that most of the shifts are valid and select shifts via AL and CIM. This example hence shows that even in settings where the exogenous share interpretation is controversial the two methods can be helpful.

\section{Conclusion}

This paper proposes adjusted shift-share IV estimators which require that only a majority or plurality of shares is valid. New statistical methods are used to select invalid shares. The STATA-programs in the supplementary material offer a simple way to apply the proposed methods.

In the migration setting, many shares are chosen as invalid and the adjusted estimates are much lower than the original ones, suggesting negative effects of immigration on wages. When including lagged migration, with the adjustments the coefficients have the expected signs. 
In this setting the proposed methods can be helpful for retrieving a causal estimate. 
In the China shock example the results are mostly robust to the use of the new methods. The results are also robust to the use of the new methods when the exclusion restriction is motivated through random shifts. 
In simulations I show that even in settings with weak instruments the estimators can continue to perform well. Severe violations of the exclusion restriction even improve the performance of the estimators in small-sample settings.

In the appendices I provide detailed descriptions of the employed methods, briefly discuss the implications of weak IVs and heterogeneous effects, provide an extension to multiple endogenous regressors and additional simulations. 

The methods are complementary to the recent literature on shift-share IVs. Before using them, it is important to think carefully about which source of validity is most feasible. The methods can be most helpful when researchers think about the shift-share instrument from the perspective of valid shares and whenever some shares are suspected to be directly correlated with the outcome variable. If there are many class-specific shocks, for example for multiple high-income countries, there is also scope for applying the methods in the quasi-random shocks setting. When doing so in this paper, my conclusions do not change, qualitatively.

I conclude with two shortcomings of the methods. First, the original methods only allow for one endogenous regressor. In the Appendix, I developed an extension of AL to multiple endogenous regressors which calls for stricter qualified majority assumptions. These new assumptions are confirmed in simulations. Further improvements would be to develop methods which can be readily extended to the multiple endogenous regressor case without making the exclusion restriction stricter. 


Second, the validity of all shares might be a concern. Given that validity of shares relies on similar arguments, it is possible that they are all inconsistent in similar ways. In this case, consistent selection can not be guaranteed. 
In fact, even though the majority and plurality assumptions are considerable relaxations of the strict exclusion restriction, they are still strict.  
Importantly, researchers should find a set of variables whose validity can be credibly defended from a theoretical point of view. The methods proposed here complement thorough theoretical considerations and do not replace them; a convincing justification of the exclusion restriction is still imperative for the new estimators.

\section*{Acknowledgements}

I'd like to thank Kirill Borusyak, David Dorn, Ben Elsner, Helmut Farbmacher, Paul Goldsmith-Pinkham, Chirok Han, Ines Helm, Stephan Huber, Peter Hull, Xiaoran Liang, Jan Stuhler, Frank Windmeijer, Joachim Winter and seminar participants at LMU Munich, TU Munich, DAGStat 2019, IAAEU, Regensburg, the Ammersee Workshop, EALE 2019 and the EALE/SOLE/AASLE World Meeting for helpful comments and discussions. I also thank Gaetano Basso and Giovanni Peri for sharing their data and code. I acknowledge funding through the International Doctoral Program ``Evidence-Based Economics'' of the Elite Network of Bavaria.

\section*{Supplementary material}
\begin{itemize}[topsep=0pt,itemsep=-1ex,partopsep=1ex,parsep=1ex]
	\item STATA-program to run Confidence Interval Method
	\item Code to reproduce results in sections \ref{sec:BP} and \ref{sec:ADH} and in Appendix \ref{sec:MCsim}.
	\item Methodological appendix with details on the methods and additional simulations
\end{itemize}



\bibliography{IVSelection}

\newpage

\newpage

\newgeometry{left=1in, top=1in,right=1in, bottom=1in}	

\section*{\Large Online Supporting Material for ``Relaxing the Exclusion Restriction in Shift-Share Instrumental Variable Estimation''}

\begin{appendices}
	\section{Methodological appendix}\label{app:MethodologicalAppendix}

\subsection{Adaptive Lasso}\label{sec:AL}

\subsubsection{Method}
I first present the AL for IV selection developed by \citet*[][WFDS]{Windmeijer2019Use}. The method consists of two parts. In short, the method chooses invalid instruments to then apply 2SLS with the instruments which have been selected as invalid. 

The method consists of three parts. 
In the first part, an initial consistent estimator $\beta_m$ is obtained through the median of 
IV estimates of exactly identified models \citep*{Han2008Detecting}. 
From this estimate $\hat{\beta}_m$, a plug-in estimate $\hat{\alpha}_m$ can be directly obtained. 
This estimator is consistent when Assumption \ref{ass:Majority} holds. 
The intuition for why the median estimate is consistent is the following: More than one half of IV-estimators which use only one IV at a time are consistent if a majority of IVs is valid. More than half of the points will hence converge to the same value. The median then will pick one of the consistent estimates. 
Why shouldn't the analysis stop here? \citet*{Windmeijer2019Use} show that even though it is consistent, the estimator has an asymptotic bias. Also, the limiting distribution is that of the order statistic of a normal distribution and this distribution is unknown, making inference on the parameter difficult. Moreover, the median estimate does not use the information contained in the additional valid IVs, missing out on efficiency gains.

In the second part, the AL uses the initial consistent estimates $\hat{\alpha}_{m,j}$ as weights. 
The AL minimization problem is
\begin{equation}\label{eq:adAL}
	\hat{\bm{\alpha}}^\lambda_{ad} = argmin \frac{1}{2} \lvert\lvert \mathbf{y} - \mathbf{\tilde{Z}}\bm{\alpha} \rvert \rvert^2_2 + \lambda_n \sum_{j=1}^J \frac{\lvert \alpha_j \rvert}{\lvert \hat{\alpha}_{m,j} \rvert} \text{,}
\end{equation}
where $\mathbf{\tilde{Z}} =\mathbf{ M_{\hat{x}}} \mathbf{Z}$, $\hat{\mathbf{x}}$ is the linear projection of $\mathbf{x}$ on the subspace orthogonal to $\mathbf{Z}$ and $\hat{\alpha}_{m,j}$ is the initial consistent estimate of $\alpha_j$, directly obtained from $\hat{\beta}_m$. For a given value of the penalty parameter $\lambda$, some entries of $\bm{\alpha}$ will be shrunk to zero. 
In the third part, IVs associated with an $\hat{\alpha}_{ad, j}$ of zero are used as valid in 2SLS estimation and those associated with non-zero coefficients are used as controls.

\noindent
In summary, the estimation procedure works as follows:
\begin{enumerate}[topsep=0pt,itemsep=-1ex,partopsep=1ex,parsep=1ex]
	\item Compile the vector of exactly identified estimates
	\item Take the median $\hat{\beta}_m$
	\item Calculate $\hat{\alpha}_m$
	\item Estimate $\alpha^\lambda_{ad}$ by adaptive Lasso
	\item 2SLS with IVs chosen as invalid included as controls and those chosen as valid used as IVs.
\end{enumerate}
The key requirement for the AL to have oracle properties is that it uses an initial consistent estimate. The key assumption for the AL to have oracle properties hence also is that the majority exclusion restriction holds.\footnote{According to Theorem 1 and Proposition 3 in WFDS, the adaptive Lasso has oracle properties when the majority exclusion restriction is fulfilled.}
As compared to the strict exclusion restriction, this assumption is already a considerable relaxation. Moreover, the AL having oracle properties does not depend on the different strength or the correlation of instruments.

For any given sample, the AL is dependent on the value of the tuning parameter $\lambda$. 
\citet*{Windmeijer2019Use} show that selection under the majority condition is consistent for any sequence of penalty parameters $\lambda_n \rightarrow \infty$ and $\lambda_n = o(n^{1/2})$, where $n$ is the number of observations. 
They propose to use the Hansen-Sargan statistic in a stopping rule, testing at each AL step, following \cites{Andrews1999Consistent} downward testing procedure. For each selection of IVs on the AL-path, the J-statistic is evaluated once. 
The authors first specify a significance level. They propose to use $0.1/ln(n)$, following \citet*{Belloni2012Sparse}. Successively smaller sets of valid IVs are tested. 
When a prespecified significance level is exceeded, the testing procedure stops. 

In applications, one might be interested in including additional endogenous regressors. However, the methods proposed here do not allow for this. Therefore, I propose a simple extension of the AL, by using an extension of the median estimator. 
In the multiple endogenous regressor case, the just-identified estimates use $P$ IVs and the estimates are stacked into matrices. In this extension I take the median along each dimension. This gives the following vector of marginal medians:

\begin{equation}
	\hat{\bm{\beta}}_m = \big( med(\tilde{\bm{\beta}}_1)\text{, ...,} med(\tilde{\bm{\beta}}_P) \big) \text{.}
\end{equation}
The key assumption for this estimator to be consistent is that the fraction of exactly identified models, which uses valid instruments exceeds 0.5. I call this the ``qualified majority condition'', because the condition on the number of valid instruments, $|\mathcal{V}|$, becomes stricter than the simple majority assumption.
\begin{assumption}\label{ass:Supermajority}
	\textbf{Qualified majority condition}
	$\frac{{|\mathcal{V}| \choose P}}{{J \choose P}}  \overset{!}{>} 0.5 \text{.}$
\end{assumption}
\noindent
For the simulations and applications, it is important to know how many valid IVs the new condition requires in the following settings. 
If we fix $J = 20$, for $P=2$ the minimum $|\mathcal{V}|$ needed to achieve an initial consistent estimate is 15 and for $P=3$ it is 17. For $P=2$ and $J=38$, it is 28. 
These assumptions are now much more strict than the simple majority condition in WFDS. A more detailed presentation of the method and discussion of the multiple regressor setting can be found in Appendix \ref{sec:Extensions}. 

\subsubsection{Illustration}
In order to illustrate the proposed methods, consider the following toy example. 
Assume that one is interested in the effect of change in immigration ($\mathbf{x}$) on change in wages ($\mathbf{y}$). The parameter of interest is $\beta$ in equation \ref{eq:ModelAugmented}. There are cross-sectional observations from $722$ commuting zones in the US. The matrix of instruments $\mathbf{Z}$ is composed of employment shares of immigrants in 1970. There are five origin countries A, B, C, D and E and hence $\mathbf{Z}$ is a $722 \times 5$ matrix. Assume that the effect of immigrants on wages is $\beta_0=0$. For countries A, B and C, the exclusion restriction is fulfilled and there is no direct effect of immigration on wages. However, the shares of countries D and E are invalid, because the base-period settlement of migrants from these countries was driven by economic factors and hence was not as-good-as-random. The selection of these five countries is the result of researcher's scrutiny, who ignores non-random settlement for countries D and E.

\begin{figure}[htb]
	\caption{Illustration\label{fig:Illustration}}
	\includegraphics[scale=0.7, trim=130 40 40 160, clip]{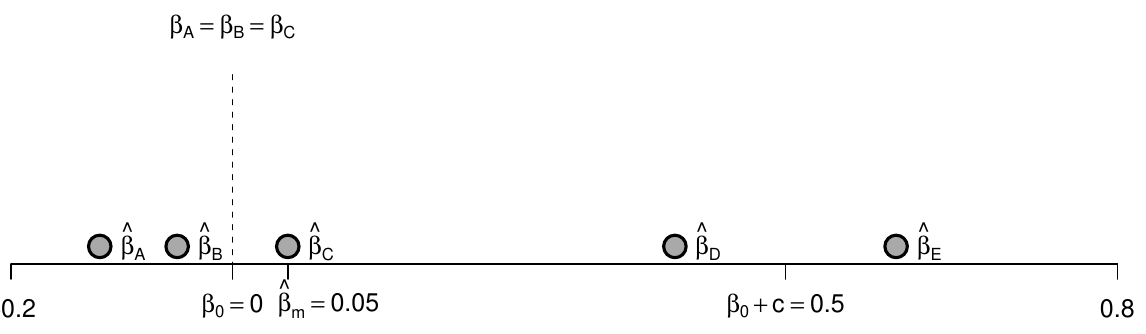}
	\floatfoot{\textit{Note:} This figure illustrates a setting with five origin country shares. 
		The values on the x-axis correspond to the values of $\hat{\beta}$. The dotted line shows the true effect $\beta_0=0$, to which the three IV estimators which use country shares A, B and C separately, converge. Estimators which use country shares D and E converge to 0.5. The grey dots illustrate the finite sample point estimates.}
\end{figure}
With AL, the first step is to use each share separately to obtain a vector of just-identified estimates. The IV estimates are illustrated in figure \ref{fig:Illustration}. The dotted, vertical line shows the true effect $\beta_0$. This effect is identified with the valid country shares A, B, and C. Let's assume that the inconsistency of the country share IV estimators D and E is $c = 0.5$. The just-identified estimates for country shares A to E are illustrated by the grey circles. In this example, the median of these estimates is $\hat{\beta}_m = \hat{\beta}_C = 0.05$. Note that if the majority exclusion is fulfilled, an IV estimator which uses a valid share is always used. From $\hat{\beta}_m$, a consistent estimate of $\bm{\alpha}$ can be obtained by
$$\hat{\bm{\alpha}}_m = (\mathbf{Z}'\mathbf{Z})^{-1}\mathbf{Z}'(\mathbf{y} - \mathbf{ M_{\hat{x}}}\hat{\beta}_m) \text{.}$$

This vector is $\hat{\bm{\alpha}}_m=(0.1, 0.2, -0.3, 1, 0.9)$. These estimates do not clearly indicate which of the shares is valid and which invalid, but they can be plugged into the AL minimization problem in equation \ref{eq:adAL}. For a specific $\lambda$, this gives us a new vector of $\bm{\alpha}$-estimates, $\hat{\bm{\alpha}}_{ad}$, where some entries are equal to zero, for example $\hat{\bm{\alpha}}_{ad}=(0,0,0,1.2,0.95)$. The vector which indicates which shares have been selected as invalid is $\hat{\mathcal{V}} = (0,0,0,1,1)$, where zero-entries denote valid shares. The first to third share vectors in $\mathbf{Z}$, are associated with a zero in the $\hat{\bm{\alpha}}_{ad}$-vector and are hence selected as valid, while the shares with non-zero values of $\hat{\bm{\alpha}}$ are chosen as invalid. These shares are finally used as instruments directly in the 2SLS, LIML or SSIV estimator.

Beginning with a very large value of $\lambda$, more importance is given to the second part of the adaptive Lasso minimization problem and no country is chosen as invalid, because all elements of $\bm{\alpha}$ are assigned zero. Adaptive Lasso is estimated via the LARS-algorithm by \citet*{Efron2004Least}, which produces a path of models, illustrated in table \ref{tab:path}. The lower $\lambda$ gets, the more shares are chosen as invalid.
\begin{table}[!htbp]
	\caption{Example of selection path \label{tab:path}}
	\begin{tabular}{ccccccc}
		\backslashbox[25mm]{country}{$\lambda=$} & $5$ &$2$ &$1.6$ &$1$ &$0.3$ & 0\\ 
		\hline
		A & 0 & 0 & 0 & 0 & 0 & 1 \\
		B & 0 & 0 & 0 & 0 & 1 & 1 \\
		C & 0 & 0 & 0 & 1 & 1 & 1 \\
		D & 0 & 1 & 1 & 1 & 1 & 1\\
		E & 0 & 0 & 1 & 1 & 1 & 1\\
		\hline
		Hansen (p) & 0.0005 & 0.001& 0.07 & 0.21 & - & - \\
		\hline
	\end{tabular}
	\floatfoot{\textit{Note:} This table illustrates an AL selection path, showing selection vector $\mathcal{V}$ for different values of penalty parameter $\lambda$.}
\end{table}
The Hansen-J over-identification test is performed for each model along this path and the corresponding p-value is compared with the pre-specified significance level of $0.1/ln(722) = 0.0152$ at each step. In this illustrative example, the Hansen-test would correctly suggest to select the oracle model in column 3, with countries A, B and C selected as valid and D and E as invalid, because for this model, the p-value of the test is larger than $0.0152$.

\subsection{Confidence Interval Method}\label{sec:CIM}

\subsubsection{Method}
Next, I present the Confidence Interval Method, which relies on an exclusion restriction which is relaxed even further. 
\citet*[WLHB, ][]{Windmeijer2020Confidence} propose the Confidence Interval Method (CIM) which builds on \cites{Guo2018Confidence} two-stage hard-thresholding. 
The idea behind the CIM is that IV estimators which use one valid instrument at a time converge to the same value. The method works as follows: 
\begin{enumerate}[topsep=0pt,itemsep=-1ex,partopsep=1ex,parsep=1ex]
	\item Set a critical value $\psi$ and calculate a confidence interval (CI) for each just-identified estimate.
	\item Confidence intervals are ordered by their lower endpoints.
	\item Lower endpoints of CIs are compared to the upper endpoints of each CI preceding it in order. If the upper endpoint of the $k$-th interval is larger than the lower endpoint of the $j$-th interval, the estimates are said to belong to the same group. The points $cil_{j}$ and $ciu_{j}$ denote lower and upper endpoints of the CI when using the $j$-th instrument. The number of overlapping intervals when comparing from instrument $j$'s CI downwards is $no_{[j]} = \sum_{k=1}^{j-1} 1(ciu_{k} > cil_{j})$.
	\item The largest group corresponds to the set of estimates with the most overlapping confidence intervals.
\end{enumerate}
\noindent 
%
Again, the result is dependent on the value of a tuning parameter. This time $\psi$ plays the role of the tuning parameter. For large values of $\psi$, all CIs will overlap and hence all variables will be chosen as valid. Gradually decreasing the value of $\psi$ narrows the confidence intervals down, and decreases the number of IVs chosen as valid. Analogously to AL, the HS test is used in a testing procedure to choose an optimal level of $\psi$. WLHB formally prove consistency of the Hansen-Sargan testing procedure.

The exclusion restriction needed for AL is stricter than the exclusion restriction needed for CIM. Why should a researcher then rely on AL? First, AL is more established than CIM. Second, the path of AL is more stable than that of CIM. 
A cautious researcher should use both methods and compare their results. If many IVs are chosen as invalid, suggesting a violation of the majority assumption, one should concentrate on the results of CIM.

\subsubsection{Illustration}\label{sec:Illustration}

\begin{figure}[!htb]
	\caption{Illustration of Confidence Interval Method\label{fig:CIM}}
	\includegraphics[scale=0.4, trim=0 20 35 60, clip]{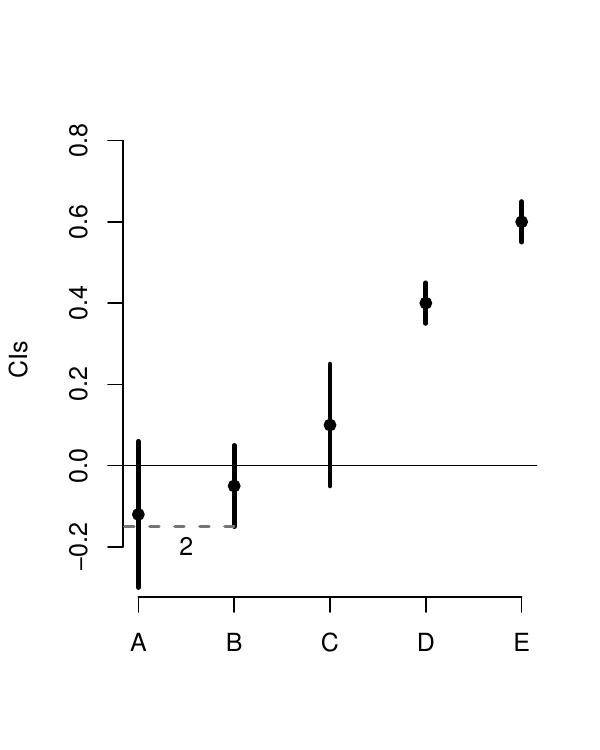}
	\includegraphics[scale=0.4, trim=42 20 35 60, clip]{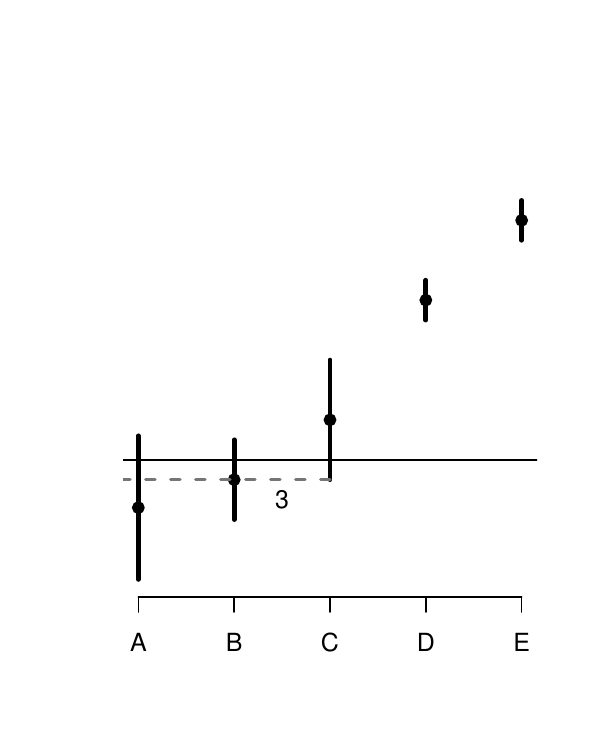}
	\includegraphics[scale=0.4, trim=42 20 35 60, clip]{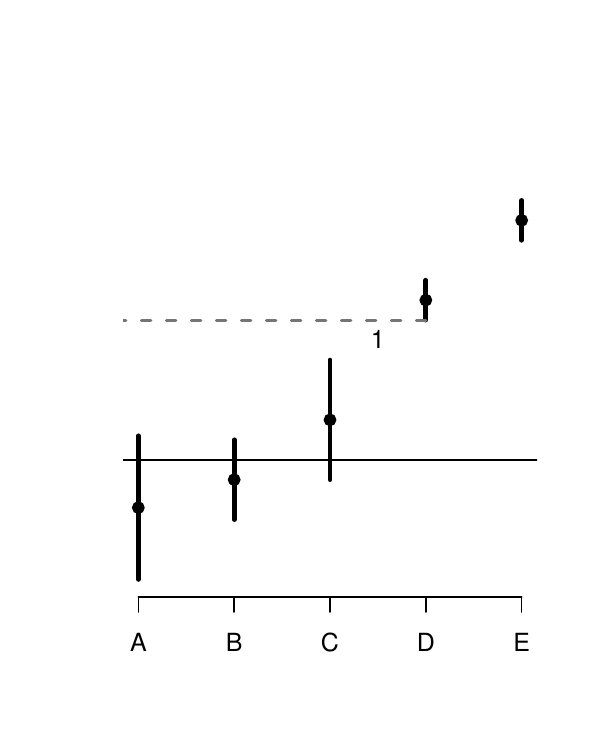}
	\includegraphics[scale=0.4, trim=42 20 35 60, clip]{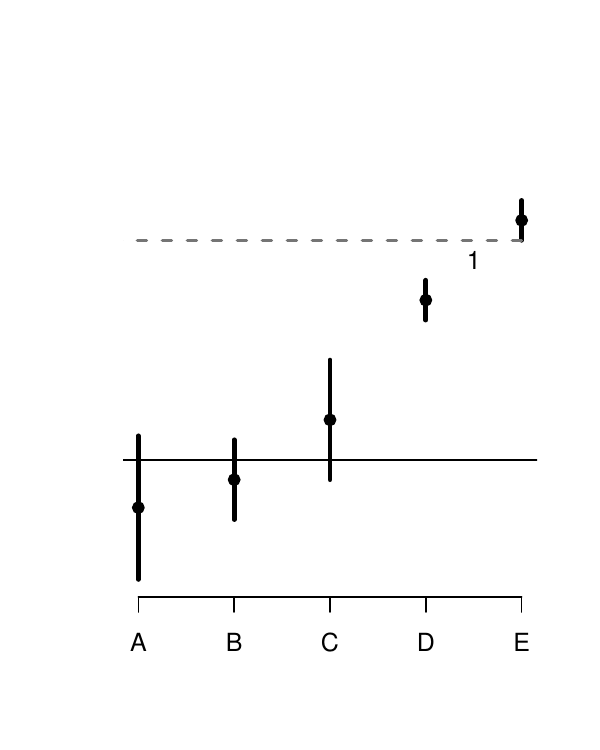}
	\floatfoot{\textit{Note:} This figure illustrates the Confidence Interval method. Black lines and dots show point estimates with intervals. Which country share is used as IV is denoted on the x-axis. Starting from the second confidence interval, the lower endpoints are compared with the upper endpoints of the CIs preceding in order. The dashed line illustrates this comparison. If the CIs overlap, they are said to belong to the same group. The number under the dashed line denotes how many IVs are in a certain group, according to the corresponding comparison. In this graph, the second panel from the left shows the largest group, which includes countries A, B, and C.}
\end{figure}

The Confidence Interval Method computes confidence intervals for each just-identified estimate and orders them by the lower endpoint of the CIs. Then it counts how often a given CI overlaps with the preceding CIs. The largest overlapping group is chosen as valid. Figure \ref{fig:CIM} illustrates the method in the toy example presented above. The second comparison, from the confidence interval of country C downwards, already selects the largest group, which includes countries A, B, and C. The other groups include only one or two IVs. In practice, the algorithm starts with a large critical value for the CIs, so that all confidence intervals overlap. Decreasing this critical value $\psi$ produces a selection path analogous to the AL selection path illustrated in table \ref{tab:path} and the algorithm stops when a prespecified level of the Hansen-J test is exceeded.

\noindent The two-stage least squares estimators adjusted by AL or CIM now estimate the following model:
$$\mathbf{y} = \mathbf{x} \beta + \mathbf{Z}_{DE} \bm{\alpha}_{DE} + \bm{\varepsilon}$$
$$\mathbf{x} =  \mathbf{Z}\bm{\gamma} + \bm{\nu}\text{,}$$
\noindent
where the shares of people from countries D and E are additionally used as controls and the rest of the country shares are used as instruments.

The illustration of the methods in figures \ref{fig:Illustration} and \ref{fig:CIM} can also be used to understand the heterogeneous effect case. Imagine that countries A, B and C still constitute the largest group, but country shares D and E are now valid IVs which estimate effects $\beta_D = \beta_E = 0.5$ consistently. The Confidence Interval Method treats countries D and E as a different causal mechanism, reporting the effect of countries A, B and C.

\subsection{Discussion: Weak instruments and heterogeneous effects}\label{sec:Discussion}
There are two main concerns with the proposed methods: weak instruments and heterogenous effects. 
In applications, weak IV bias is a concern, because each share is used individually to predict the endogenous variable. 
There are three answers to this problem:
First, the limited information maximum likelihood (LIML) which has better finite-sample properties than the two-stage least squares can be used after selection. Second, 
I also use the Anderson-Rubin test statistic in the downward testing procedure to use a test which can detect violations of the exclusion restriction in presence of weak instruments. Third, in simulations I show that the algorithms also perform well, when IVs are weak. 
Still, these are just practical answers to weak instruments. How to address weak instruments in valid IV selection is the object of future research. 


The methods presented in this paper rely on the constant treatment effect assumption. 
\citet*{Goldsmith-Pinkham2020Bartik} allow for location-specific coefficients $\beta_l$. When all the first-stage coefficients have the same sign (monotonicity), each class-specific IV estimates a weighted average of location-specific effects,
\begin{equation}\label{eq:LATE}
	\underset{L \rightarrow \infty}{plim}
	\hat{\beta}_{j} = E(\omega_{lj}\beta_l) \text{,}
\end{equation}
according to Proposition 4.1 in \citet*{Goldsmith-Pinkham2020Bartik}. The definition of $\omega_{lj}$ can be found there. 
In a LATE-framework the shift-share estimator is therefore a weighted combination of class-specific weighted averages. 
These class-specific estimates $\hat{\beta}_j$ therefore might differ. 

Can the selection methods proposed before deal with heterogeneous effects? One practical solution is to perform the analogous analyses for clusters of industries, inside of which the constant treatment effect assumption is believed to hold. If the constant effect assumption is more generally violated, another approach is needed. Combining heterogeneous treatment effects with valid IV selection is also the object of current research and an exhaustive answer is beyond the scope of this paper. 

	\subsection{Additional notation}
	The $n \times n$ projection matrix is $\mathbf{P_X} \equiv \mathbf{X}(\mathbf{X}'\mathbf{X})^{-1}\mathbf{X}'$, the annihilator matrix is $\mathbf{M_X} = \mathbf{I} - \mathbf{P_X}$, $\mathbf{\hat{x}} = \mathbf{P_Z} \mathbf{x}$ are the fitted values from running a regression of the endogenous regressors on the instruments and $\tilde{\mathbf{Z}} \equiv \mathbf{M_{\hat{x}}} \mathbf{Z}$.
	
	\subsection{Oracle estimator}\label{app:Oracle}
	In this subsection I introduce the oracle model as defined in \citet*{Windmeijer2020Confidence}.
	$$\mathbf{y} = \mathbf{x}\beta + \mathbf{Z}_{\mathcal{I}}\bm{\alpha}_{\mathcal{I}} + \mathbf{u}$$
	where $\mathcal{I}$ denotes the true set of invalid IVs. The oracle estimator then is 
	$$\hat{\beta}_{or} = \left(\hat{\mathbf{x}}' \mathbf{M}_\mathcal{I} \hat{\mathbf{x}} \right)^{-1} \hat{\mathbf{x}}' \mathbf{M}_\mathcal{I} \mathbf{y}\text{,}$$
	where $\mathbf{M}_\mathcal{I} = \mathbf{M}_{Z_\mathcal{I}}$. Under Assumptions \ref{ass:FirstStage}, \ref{ass:Rank}, \ref{ass:Asymptotics} (below) and as $n \rightarrow \infty$
	
	\begin{equation*}
	\sqrt{n} \left(\hat{\beta}_{or} - \beta\right) \overset{d}{\rightarrow} N\left(0,\sigma_{or}^2\right)\text{,}
	\end{equation*}
	where $\sigma_{or}^2 = \sigma_u^2 \left(plim(\frac{1}{n} \hat{\mathbf{x}}' \mathbf{M}_\mathcal{I} \hat{\mathbf{x}})^{-1}\right)$.
	
	\subsection{Details on adaptive Lasso with multiple endogenous regressors}\label{sec:Extensions}
	In this Appendix, I provide additional details on the AL as presented in section \ref{sec:AL}. 

	\subsubsection{Model and assumptions}

	With multiple endogenous regressors, the first stages are
	\begin{equation}\label{eq:FirstStage-Appendix}
	\mathbf{X_p} = \mathbf{Z} \gamma_p + \bm{\varepsilon}_p, \text{ for } p \in \{1, ..., P\} \text{.}
	\end{equation}
	There are now $P$ endogenous regressors $x_1$, ..., $x_P$, which can be subsumed in a matrix $\mathbf{X}$, $J$ instrument vectors $\mathbf{z}_1$, ..., $\mathbf{z}_J$, which can be subsumed to a matrix $\mathbf{Z}$ and error terms $\mathbf{u}$ and $\bm{\varepsilon}_p$ for $p \in \{1, ..., P\}$ which can be correlated with cov$(\mathbf{u}, \bm{\varepsilon}_P) \equiv \rho_P$. The latter covariance measures the endogeneity of regressors in $\mathbf{x}$. The coefficient vector of interest is $\bm{\beta}$ ($P \times 1$). The $(J \times 1)$-vector of first-stage coefficients for regressor $p$ is $\bm{\gamma}_p$. 
	Define $g \equiv |\mathcal{V}|$ as the number of valid instruments, i.e. the instruments for which $\alpha_j = 0$. If we estimate all possible combinations of exactly identified models, taking $P$ out of $J$ instruments at a time, we get ${J \choose P}$ estimate vectors. 
	Let $\mathbf{Z}^{[j]}$ with $j \in \{1, ..., {J \choose P}\}$ be the $n \times P$ matrix of one combination of instruments.
	
	Considering the model from the main text (equations \ref{eq:ModelAugmented} and \ref{eq:ModelFS}) and following WFDS, I assume the following:
	
	\setcounter{assumption}{5}
	\begin{assumption}\label{ass:FirstStage} \textbf{Existence of first stages}
		\\ For all possible values of $[j]$, let $\bm{\gamma}_{[j]}$ be the combinations of the $k_{th}$-row of $\bm{\gamma}$ for all $k \in [j]$. Then assume  $$rank(\bm{\gamma}_{[j]})=P$$
	\end{assumption}
	
	\begin{assumption}\label{ass:Rank}
		\textbf{Rank condition}\\
		\begin{equation*}
		rank(E[\mathbf{Z'}\mathbf{Z}]) = J
		\end{equation*}
	\end{assumption}
	
	\begin{assumption}\label{ass:Errors} \textbf{Error structure}\\
		$$Var(\bm{u}) = \sigma_u^2, \quad Var(\bm{\varepsilon}_p) = \sigma_{\bm{\varepsilon}_p}^2 \quad Cov(\mathbf{u}, \bm{\varepsilon}_p) = \sigma_{u, \varepsilon_{p}} \quad \mathbf{w} = (\mathbf{u} \quad \bm{\varepsilon})^\intercal$$ 
		$$ \text{e.g. for } P=2 \quad E[\bm{w}^\intercal \bm{w}]= \left( \begin{array}{rrr}
		\sigma_u^2 & \sigma_{u, \varepsilon_{1}} & \sigma_{u, \varepsilon_{2}} \\
		\sigma_{u, \varepsilon_{1}} & \sigma_{\varepsilon_1}^2 & \sigma_{\varepsilon_{1}, \varepsilon_{2}} \\
		\sigma_{u, \varepsilon_{2}} & \sigma_{\varepsilon_{1}, \varepsilon_{2}} & \sigma_{\varepsilon_2}^2\\
		\end{array}\right)
		\equiv \bm{\Sigma} $$
	\end{assumption}
	
	\begin{assumption}\label{ass:Asymptotics}
		\begin{align}
		plim (n^{-1} \mathbf{Z}^\intercal \mathbf{Z}) &= E(\mathbf{Z}^\intercal \mathbf{Z})
		\quad ; \quad plim(n^{-1} \mathbf{Z}^\intercal \mathbf{X}) = E(\mathbf{Z}^\intercal \mathbf{X}) \notag\\
		plim(n^{-1} \mathbf{Z}^\intercal \mathbf{u})&= E(\mathbf{Z}^\intercal \mathbf{u}) = 0\notag\\ 
		plim(n^{-1} \mathbf{Z}^\intercal \bm{\varepsilon_p}) &= E(\mathbf{Z}^\intercal \bm{\varepsilon_p}) = 0\notag\\
		plim(n^{-1}\sum\limits_{i=1}^{n} w_i) &= 0 \quad ; \quad plim(n^{-1} \mathbf{w}^\intercal \mathbf{w}) = \bm{\Sigma} \notag
		\end{align}
	\end{assumption}
	\noindent Assumptions \ref{ass:Supermajority},  \ref{ass:FirstStage} and  \ref{ass:Rank} imply Assumption 3 in \citet{Han2008Detecting} (with the difference that in our example we abstracted from covariates).	
	Assumptions \ref{ass:Rank}, \ref{ass:Errors} and \ref{ass:Asymptotics} ensure the existence of the Hansen-Sargan test statistic. 	
	Assumption \ref{ass:FirstStage}, is the standard relevance assumption for each just-identified model. This assumption implies, that the first stage coefficients are all non-zero and is crucial for identification.
	
	As will be shown below, the vector of marginal medians of the matrix of estimates from exactly identified models is a consistent estimator of $\bm{\beta}$. This proof follows closely the one implied by \citet{Han2008Detecting}.

	\subsubsection{Additional examples of the qualified majority}
	For the two-regressor case, P = 2, it follows:
	\begin{corollary}
		When $P=2$ and the number of candidate instruments grows to infinity, the fraction of valid IVs has to exceed 70,7107\%.
	\end{corollary}
	
	\textbf{Proof:} \begin{equation}\label{eq:Corollaryk2}
	\frac{\frac{g!}{2!(g-2)!}}{\frac{J!}{2!(J-2)!}} = \frac{(g-1)g}{(J-1)J} \overset{!}{>} 0.5 \quad 
	\Rightarrow \quad g \overset{!}{>} \frac{1 + \sqrt{1 + 4\cdot 0.5 \cdot (J-1) \cdot J}}{2}\text{.}
	\end{equation}
	Multiply both sides in \ref{eq:Corollaryk2} with $1/J$ and take $\Lim{J \rightarrow \infty}$. $\qed$ 
	
\begin{center}
		\begin{figure}[!ht]
			\includegraphics[scale=0.6]{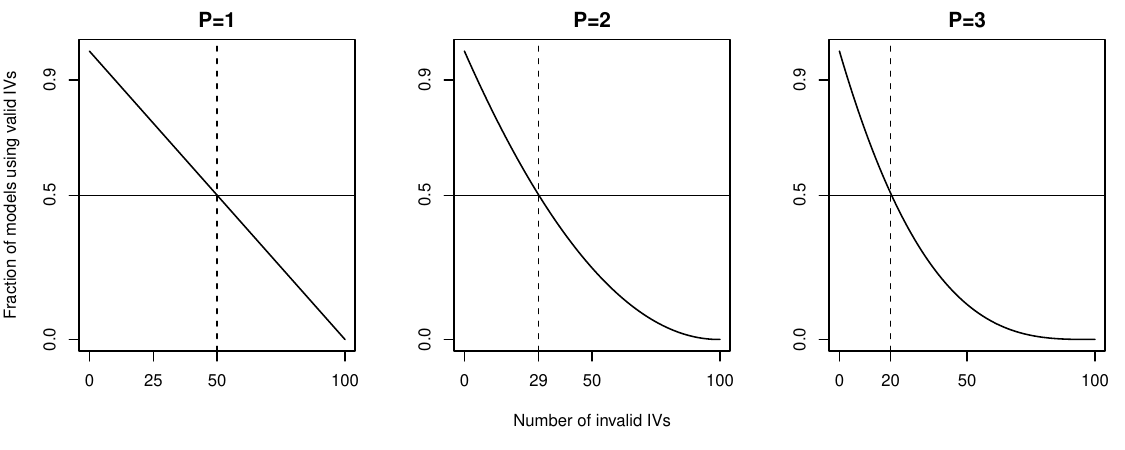}	
			\caption{Fraction of valid instruments against number of valid instruments. In the left panel $P=1$, in the middle panel $P=2$ and in the right panel $P=3$. \label{fig:CutL100}}
		\end{figure}
	\end{center}

	To get a better feeling about how the majority condition is altered with the number of endogenous regressors in the model, we fix $J$ at 100. The minimal $g$ needed for oracle properties is found by plugging into \ref{ass:Supermajority} and sequentially decreasing $g$ until the condition is fulfilled. 
	For $P=3$ the fraction of valid instruments needs to be at least 80, for $P=4$, 85; for $P=5$, 88; for $P=6$, 90; for $P=15$, 96 and for $P=30$, 99. The relationship between number of invalid IVs and fraction of models using only valid IVs is visualized in \ref{fig:CutL100}. 
	
	With growing number of endogenous regressors, the number of exogenous instruments needed also grows. In the limiting case, when the number of endogenous regressors is maximal, it is equal to the number of candidate instruments. In this case, we have only one exactly identified model. The method does not provide any benefit then, because it cannot discard any instrument, as the model would then be underidentified. Assumption \ref{ass:Supermajority} now becomes a consensus rule: all of the instruments need to be valid. 
	
	\subsubsection{Consistency of the vector of marginal medians}
	I rewrite estimators as in the proof of Proposition A1 in \citet*{Windmeijer2020Confidence}. First, partition the matrix $\mathbf{Z} = (\mathbf{Z_1} \quad \mathbf{Z_2})$, where $\mathbf{Z_1}$ is a $n \times P$ and $\mathbf{Z_2}$ is a $n \times (J-p)$ matrix. $\bm{\gamma} = (\bm{\gamma_1} \quad \bm{\gamma_2})'$ is the equivalent partition of the matrix of first-stage coefficients. $\mathbf{Z^*} = [\mathbf{\hat{X}\quad \mathbf{Z_2}}]$, then $\mathbf{Z^*} = \mathbf{Z} \mathbf{\hat{H}}$, with
	
	\[
	\hat{\mathbf{H}} =
	\left( {\begin{array}{cc}
		\bm{\hat{\gamma}_1} & 0 \\
		\bm{\hat{\gamma}_2} & \mathbf{I}_{J - P}\\
		\end{array} } \right)
	; \quad 
	\hat{\mathbf{H}}^{-1} =
	\left( {\begin{array}{cc}
		\bm{\hat{\gamma}_1^{-1}} & 0 \\
		-\bm{\hat{\gamma}_2\hat{\gamma}_1^{-1}} & \mathbf{I}_{J - P}\\
		\end{array} } \right)
	\]
	Each of the $\binom{J}{P}$ estimators can be written as 
	\[
	(\hat{\bm{\beta}}_{2SLS} \quad \hat{\bm{\pi}}_{2SLS})' =  \mathbf{\hat{H}^{-1}}\bm{\hat{\Gamma}} = \mathbf{\hat{H}^{-1}}(\mathbf{Z}^\intercal \mathbf{Z})^{-1} \mathbf{Z}^\intercal (\mathbf{X}\bm{\beta} + \mathbf{Z} \bm{\alpha} + \mathbf{u}) = \mathbf{\hat{H}^{-1}}(\bm{\hat{\gamma}}\bm{\beta} + \bm{\alpha})
	\]
Note that the part in parentheses is equal to 	
	\[
	\left(
	\begin{array}{c}
	\bm{\gamma_1 \beta + \alpha_1}\\
	\bm{\gamma_2 \beta + \alpha_2}\\
	\end{array}
	\right) \text{.}
	\]	
	The resulting $J \times 1$ vector will be 
	\[
	(\bm{\hat{\beta}_{2SLS}} \quad \hat{\bm{\pi}}_{2SLS})'
	= \left(
	\begin{array}{c}
	\bm{\beta + \gamma_1^{-1}\alpha_1}\\
	\bm{- \gamma_2 \gamma_1^{-1}\alpha_1 + \alpha_2}
	\end{array}
	\right)
	\]
	Hence, the inconsistency of $\bm{\beta}$ is $\bm{c} \equiv \bm{\hat{\beta}_{2SLS}} - \bm{\beta} = \bm{\gamma_1}^{-1}\bm{\alpha_1}$.
	
	$\tilde{\bm{\beta}}$ is the ${J \choose P} \times P$ matrix, which consists of the stacked $\bm{\hat{\beta}_{2SLS}}$ row vectors. 
	The easiest way to compute a median for this multidimensional set of points is to take the median along each of the $P$ dimensions separately.\footnote{I have considered alternatives to the vector of marginal medians, such as the Tukey median or Oja's simplex median, but have not implemented them because they are not continuous functions.} The resulting $P$-dimensional vector is called the vector of marginal medians:
	\begin{equation}
	\bm{\hat{\beta}_m} = \big( med(\bm{\tilde{\beta}_1})\text{, ...,} med(\bm{\tilde{\beta}_P}) \big) \text{.}
	\end{equation}
	
	\begin{proposition}
		Under assumptions \ref{ass:Supermajority}, \ref{ass:FirstStage} and \ref{ass:Rank}, $\bm{\hat{\beta}_m} \overset{P}{\rightarrow} \bm{\beta_0}$, where $\bm{\beta_0}$ is the true $P \times 1$ $\bm{\beta}$-vector.
	\end{proposition}
	
	\textbf{Proof:} 
	Let $\tilde{\beta}^{[j]}_{p}$ be the 2SLS-estimator for the $p$-th coefficient $\beta_p$ when a certain combination $[j]$ of IVs is used.
	Then $$plim(\tilde{\beta}^{[j]}_{p}) = \beta_0 + c^{[j]}_{p} \quad \forall \quad j$$ 
	hence $$plim\{(\bm{\tilde{\beta}_{1}}, \bm{\tilde{\beta}_{2}},..., \bm{\tilde{\beta}_{P}})\} = (\bm{\beta_0} + \bm{c_{1}}, \bm{\beta_0} + \bm{c_{2}}, ..., \bm{\beta_0} + \bm{c_{P}})$$
	where $c^{[j]}_{p}$ is the $p$-th entry of  $\bm{c^{[j]}}=\bm{\gamma_1}^{-1}\bm{\alpha_1}$, i.e. the inconsistency from using at least one invalid instrument in the $[j]$-set. There are $\binom{J}{P}$ entries in $\mathbf{c_p}$, which is the vector collecting inconsistency terms for a certain regressor, when using all possible combinations of IVs.
	
	The median function is continuous. Hence, by the continuous mapping theorem (CMT): $$plim\{med(\bm{\tilde{\beta}_{1}}), med(\bm{\tilde{\beta}_{2}}),..., med(\bm{\tilde{\beta}_{P}})\} = \{med(\bm{\beta_{0,1}} + \bm{c_{1}}), ..., med(\bm{\beta_{0,P}} + \bm{c_{P}})\}$$
	Under assumption \ref{ass:Supermajority}, $c^{[j]}_{p} = 0$ holds for a majority of entries inside each column.
	Then $$plim\{\bm{\hat{\beta}_m}\} = \bm{\beta_0} {.} \qed$$
	
	\subsubsection{Adaptive Lasso}
	
	Given the consistent estimator $\bm{\beta}_m$, the following procedure is analogous to WFDS. For a detailed overview over the original AL method, refer to WFDS and \citet{Zou2006adaptive}. A consistent estimator for $\bm{\alpha}$ can be achieved by rewriting the moment conditions $E(\mathbf{Z}'\mathbf{u})=0$:
	
	\begin{equation}
	\hat{\bm{\alpha}}_m = (\mathbf{Z}^\intercal \mathbf{Z})^{-1} \mathbf{Z}^\intercal (\mathbf{y} - \mathbf{X}\hat{\bm{\beta}}_m)\text{.}
	\end{equation}
	The adaptive Lasso (AL) estimator as proposed in \citet{Zou2006adaptive} can be used, where the penalty term is weighed by the initial consistent estimate: 
	
	\begin{equation}\label{eq:AL}
	\hat{\bm{\alpha}}^\lambda_{ad} = \argmin_{\bm{\alpha}} \frac{1}{2} \lvert\lvert \mathbf{y} - \mathbf{\tilde{Z}}\bm{\alpha} \rvert \rvert^2_2 + \lambda_n \sum_{j=1}^J \frac{\lvert \alpha_l \rvert}{\lvert \hat{\alpha}_{m, j}\rvert} \text{.}
	\end{equation}
	
	\noindent The AL estimator $\bm{\beta}_{ad}$ is then retrieved from the conditions $E(\hat{\mathbf{D}}'\mathbf{u}) = E\big(\hat{\mathbf{D}}'(\mathbf{y} - \hat{\mathbf{Z}}\bm{\alpha} - \mathbf{\hat{D}} \bm{\beta})\big) = 0$
	
	\begin{equation}\label{eq:adaBeta}
	\bm{\beta}^\lambda_{ad} = \big(\hat{\mathbf{X}}^\intercal\mathbf{\hat{X}}\big)^{-1} \hat{\mathbf{X}}(\mathbf{y} - \mathbf{Z}\hat{\bm{\alpha}}^\lambda_{ad})\text{.}
	\end{equation}

	\section{Monte Carlo simulations}\label{sec:MCsim}
	
	The following simulations illustrate that AL and CIM select the invalid shares as invalid, when the sample is reasonably large. The adjusted estimators perform better in terms of bias as compared to the standard 2SLS estimator, which uses all the shares, irrespective of their validity. Moreover, I show that even with weak instruments good selection results are possible and that strong violations of the exclusion restriction even improve their performance. Finally, I show that with multiple regressors a higher fraction of IVs needs to be valid for the AL to have oracle properties.
	
	\subsection{Single regressor}
	
	\subsubsection{Setup}
	
	The data is created based on the model in \ref{eq:ModelAugmented} and \ref{eq:ModelFS}. 
	The coefficient of interest $\beta$ is set to 0 and the elements of the first-stage coefficient vector $\gamma$ to 0.6. 
	To create ten shares, I draw a matrix with $J=10$ columns from a uniform distribution between 0 and 0.1. The vector of direct effects of the IVs, $\bm{\alpha}$, is set to $\bm{\alpha} = (0.2, 0.2, 0.2, 0, 0, 0, 0, 0, 0, 0)$, so that a majority of shift-share products is still valid. In a second simulation, the vector is set to $\bm{\alpha} = (0.1, 0.1, 0.2, 0.2, 0.3, 0.3, 0, 0, 0, 0)$, such that only the largest group of IVs is valid. 
	The error terms $u_l$ and $\varepsilon_l$ are distributed as 
	\begin{equation}
		\begin{pmatrix}
			u_l \\
			\varepsilon_l 
		\end{pmatrix} 
		\sim N\bigg( 
		\begin{pmatrix}
			0\\
			0
		\end{pmatrix}, \quad
		\begin{pmatrix}
			1& 0.5\\
			0.5& 1\\
		\end{pmatrix}
		\bigg)\text{.}
	\end{equation}
	I vary the sample size from 400 to 6000 observations, gradually increasing by 400. The number of repetitions is 100 for each parameter combination, each time drawing the errors anew. 
	
	\subsubsection{Results}
	
	\begin{figure}[!htb]
		\centering
		\includegraphics[scale=0.6, trim=0 120 0 0, clip]{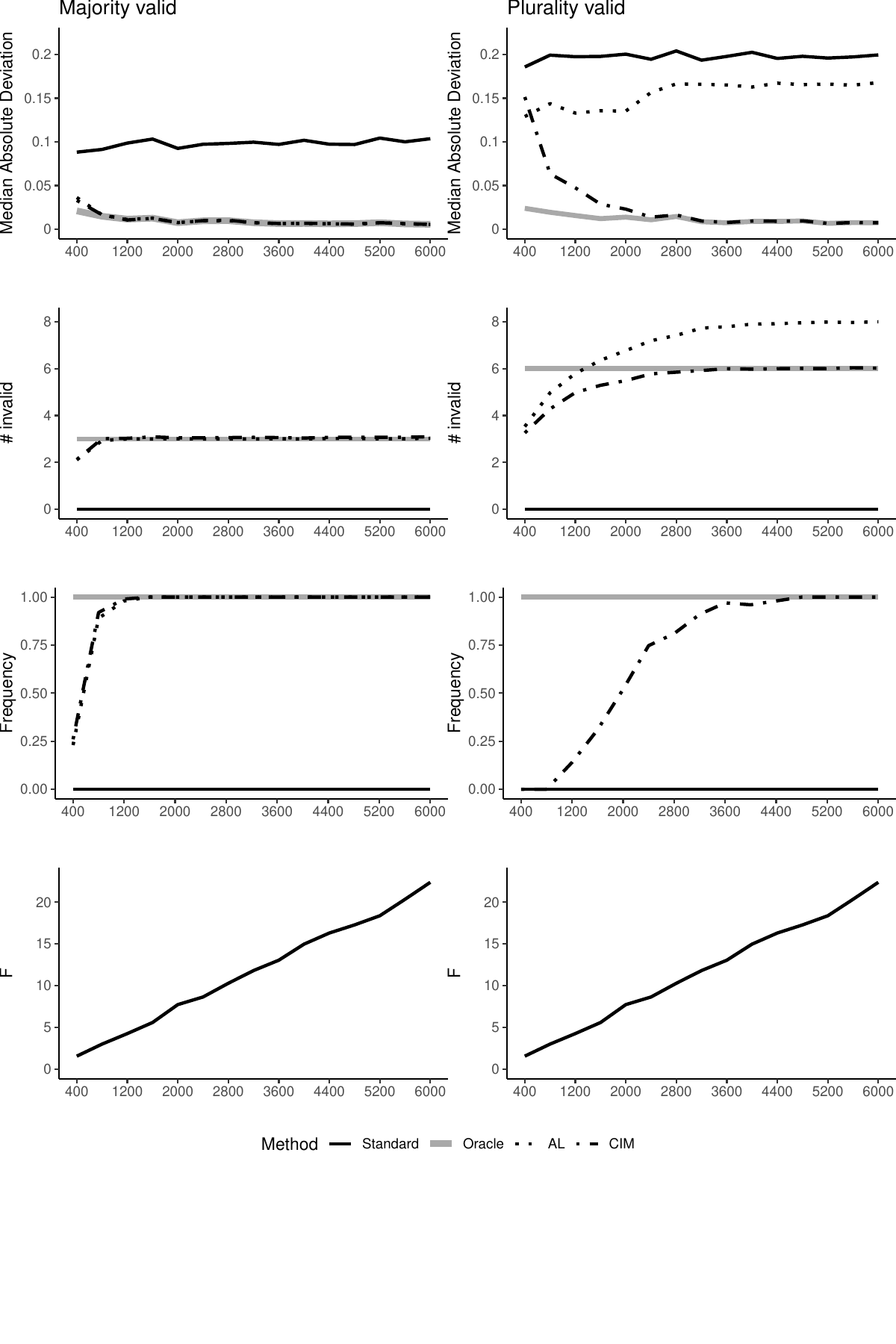}
		\caption{Simulation results}
		\floatfoot{\textit{Note:} Performance of 2SLS in Monte Carlo simulations as described in section \ref{sec:MCsim} using shares as IVs, adjusted with AL and confidence interval method. 100 replications have been used for each sample size. IVs chosen as invalid are included as controls. Horizontal axis: Number of observations. First row: median absolute deviation, second row: number of IVs chosen as invalid, third row: relative frequency with which all invalid IVs have been chosen as invalid. Grey line: oracle 2SLS, black line: standard 2SLS, dotted line: AL, dashed-dotted line: Confidence Interval Method.\label{fig:SimulRes-Main}}
	\end{figure}
	
	The two baseline estimators are the standard and the oracle 2SLS. The standard 2SLS is the estimator for which all shares are assumed to be valid, and the oracle shift-share estimator is the ideal estimator for which only valid shares are used as IVs and invalid ones are used as controls. 
	I compare these two estimators with the 2SLS estimator adjusted by AL and CIM. 
	I report the median absolute deviation (MAD), the mean number of IVs selected as invalid, the frequency with which all invalid IVs have been selected as invalid and the F-statistic of the oracle model.
	
	The main result is that the adjusted 2SLS estimators outperform the standard 2SLS estimator in the majority setting for all sample sizes and approach the performance of the ideal estimator with growing sample size. In the plurality setting only the CIM approaches the performance of the oracle estimator. 
	This is in line with the predictions: When the majority rule holds, both methods should work well and AL is expected to fail when only the plurality rule holds but majority does not.
	
	The graphs on the left of Figure \ref{fig:SimulRes-Main} depict the setting in which a majority (seven out of ten) of shares is valid. 
	The MAD of the standard estimator is at about 0.1 and does not decrease as the sample size gets larger. 
	The oracle 2SLS estimator's median absolute deviation is below 0.05 and gets closer to zero as $n$ increases. Notably, the MAD of the 2SLS adjusted by AL visualized by the dotted line and the 2SLS estimator adjusted by CIM, visualized by the dashed-dotted line, have a MAD equal to that of the oracle estimator (the grey, solid line) already for a moderate sample size of $N \geq 1000$. 
	In the second and third rows, it becomes clear why that is the case. From a sample size of 1000 upwards, only three shares are chosen as invalid on average and in 100\% of the cases the chosen IVs are the invalid ones, as can be seen in the third row. Whenever one appeals to the asymptotic properties of an estimator, the question of how large the sample size needs to be is legitimate. In the setting of this simulation, the adjusted estimators already attain oracle properties from a relatively low sample size on. 
	
	The graphs on the right of Figure \ref{fig:SimulRes-Main} show the results from the setting in which only a plurality of shares is valid (four out of ten). 
	Here, the estimator adjusted by AL fares only slightly better than the standard estimator, with a MAD at about 0.15, which does not monotonously decrease with growing sample size. On average, about eight shares are selected as invalid, as the sample gets large, but in no MC replication all invalid IVs are correctly selected as invalid. This can be seen from the right graph in row three: the dotted line and the solid black line coincide. Selection via the CIM achieves a performance which is equal to the oracle 2SLS from a sample size of about 2000 on. The average number of IVs selected as invalid reaches six when $n=2000$ and the frequency with which all invalid are selected as invalid is close to one at $n=3000$.
	
	\subsection{Weak IVs and strong violations}
	Next, I ask how weak instruments and stronger direct effects change the performance of the estimators. 
	Notably, in the lowest graphs in figure \ref{fig:SimulRes-Main} the F-statistic grows steadily with sample size but it is lower than 10 for many sample sizes. Still, even when F is lower than 5, the performance of AL and CIM is close to that of the oracle estimator. This suggests that there are settings in which weak IVs do not severely affect the selection performance. 
	
	For the additional simulations in figures \ref{fig:All-maj} and \ref{fig:All-plu}, I use the same setup as before but I change the first-stage parameter $\gamma$ and the parameter of direct effect coefficients $\alpha$. The following table explains the changes. Lines 1 and 5 replicate the original setup, already reported in table \ref{fig:SimulRes-Main}.
	\vspace{10px}
	
	\begin{center}
	\begin{tabular}{lcc}
		\hline
		Setting: & $\bm{\gamma}$ & $\bm{\alpha}$ \\
		\hline
		Majority & & \\
		\hline
		Strong IV, weak violation & $0.6$ & $0.2$ \\	
		Weak IV, weak violation & $0.3$ & $0.2$ \\
		Strong IV, strong violation & $0.6$ & $0.4$ \\	
		Weak IV, strong violation & $0.3$ & $0.4$ \\
		\hline
		Plurality & & \\
		\hline
		Strong IV, weak violation & $0.6$ & $0.1, 0.1, 0.2, 0.2, 0.3, 0.3$ \\
		Weak IV, weak violation & $0.3$ & $0.1, 0.1, 0.2, 0.2, 0.3, 0.3$ \\
		Strong IV, strong violation & $0.6$ & $0.2, 0.2, 0.4, 0.4, 0.6, 0.6$ \\	
		Weak IV, strong violation & $0.3$ & $0.2, 0.2, 0.4, 0.4, 0.6, 0.6$ \\
		\hline
	\end{tabular}	
\end{center}

\noindent The results for these simulations can be found in Figures \ref{fig:All-maj} and \ref{fig:All-plu}. In a nutshell, the results are that lowering the maximal F statistic in the simulations changes the results by little and increasing the entries of $\bm{\alpha}$ to make the direct effect stronger even improves performance in small-sample settings.
	
	\subsection{Multiple regressors}\label{sec:Multi}
	In a further simulation, I also compare the performance of the adaptive Lasso with the vector of marginal medians as an initial estimator, the standard and the oracle 2SLS estimators, but now with multiple endogenous regressors. I use 20 IVs in total and gradually increase the number of invalid instruments from zero to 18 or 17 (when $P=3$). I expect that the performance of the AL breaks down at the predicted cutoffs, when there are more than ten ($P=1$), five ($P=2$) and three ($P=3$) invalid IVs. 
	
	In the simulations involving multiple endogenous regressors, the following is equal in all settings: 
	There are 100 iterations per number of invalid IVs, the sample size is $n=10,000$, there are $L=20$ IVs. 
	The coefficient of interest $\beta$ is set to 0. $\mathbf{Z}$ is a $N \times 20$ matrix drawn from a uniform distribution between 0 and 1. The vector indicating invalidity, $\bm{\alpha}$ in equation \ref{eq:ModelAugmented}, is set to $\bm{\alpha} = (1, 0, ..., 0, 0)$, when one IV is invalid, to $\bm{\alpha} = (1, 2, 0, ..., 0, 0)$ when two IVs are invalid, etc.
	
	The error term from the structural equation is normally distributed with $\mathbf{u} \sim N(0,0.25)$. The first-stage error terms $\bm{\varepsilon}_p$ are constructed as $\bm{\varepsilon}_p = N(0,1) + 0.5 \mathbf{u}$, obtaining the variances $Var(\bm{\varepsilon}_p) = 1.0625$, and the covariances $Cov(\mathbf{u}, \bm{\varepsilon}_p) = 0.125$.
	
	For the case with one regressor, the first-stage coefficient vector $\bm{\gamma}$ is a matrix of ones only. When there are two endogenous regressors, the matrix of coefficients is made of two vectors. The first-stage coefficient vectors for the first and second regressors are $\bm{\gamma}_1 = (0.05, 0.1, 0.15, ..., 0.95, 1)$ and $\bm{\gamma}_2 = (1, 0.95, 0.85, ..., 0.1, 0.05)$. For $P=3$, the matrix is composed of the two just-mentioned vectors and of an additional one: $\bm{\gamma}_3 = (0.05, 0.1, 0.15, 0.2, ..., 0.05, 0.1, 0.15, 0.2)$.
	
\begin{figure}[htb]
		\begin{center}
			\includegraphics[scale=0.65, trim=0 20 0 0, clip]{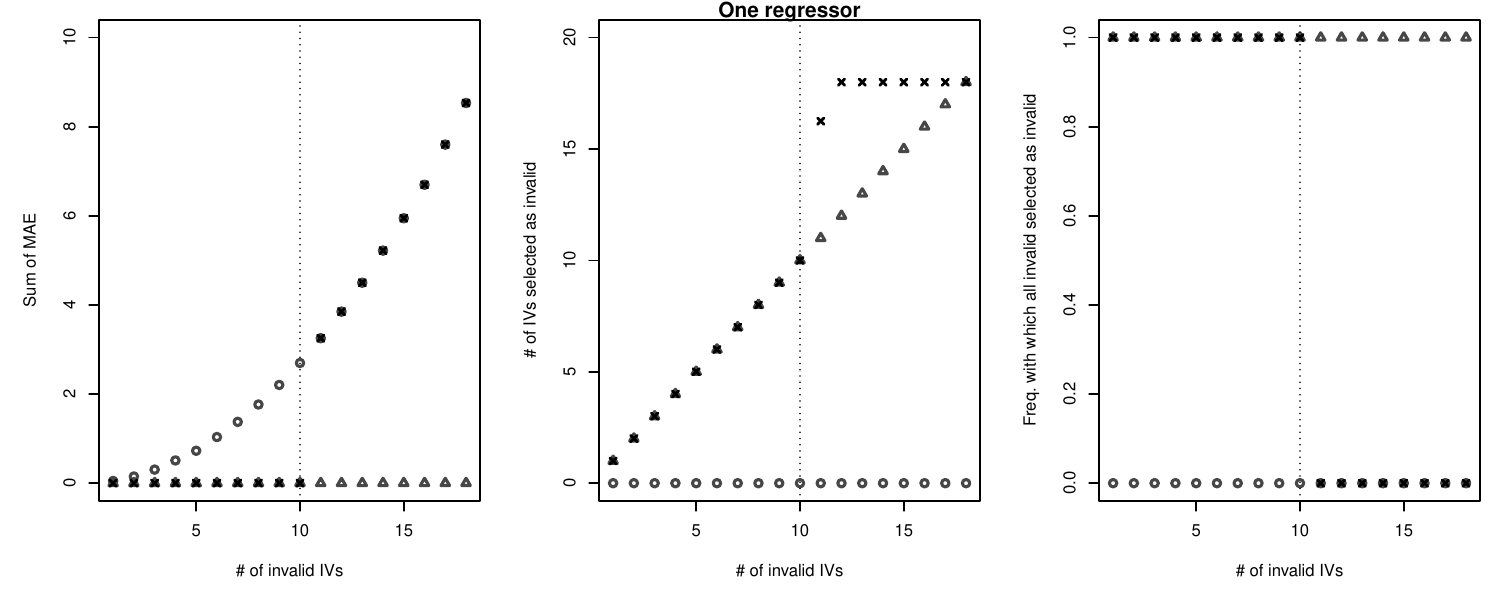}
			\includegraphics[scale=0.65, trim=0 20 0 0, clip]{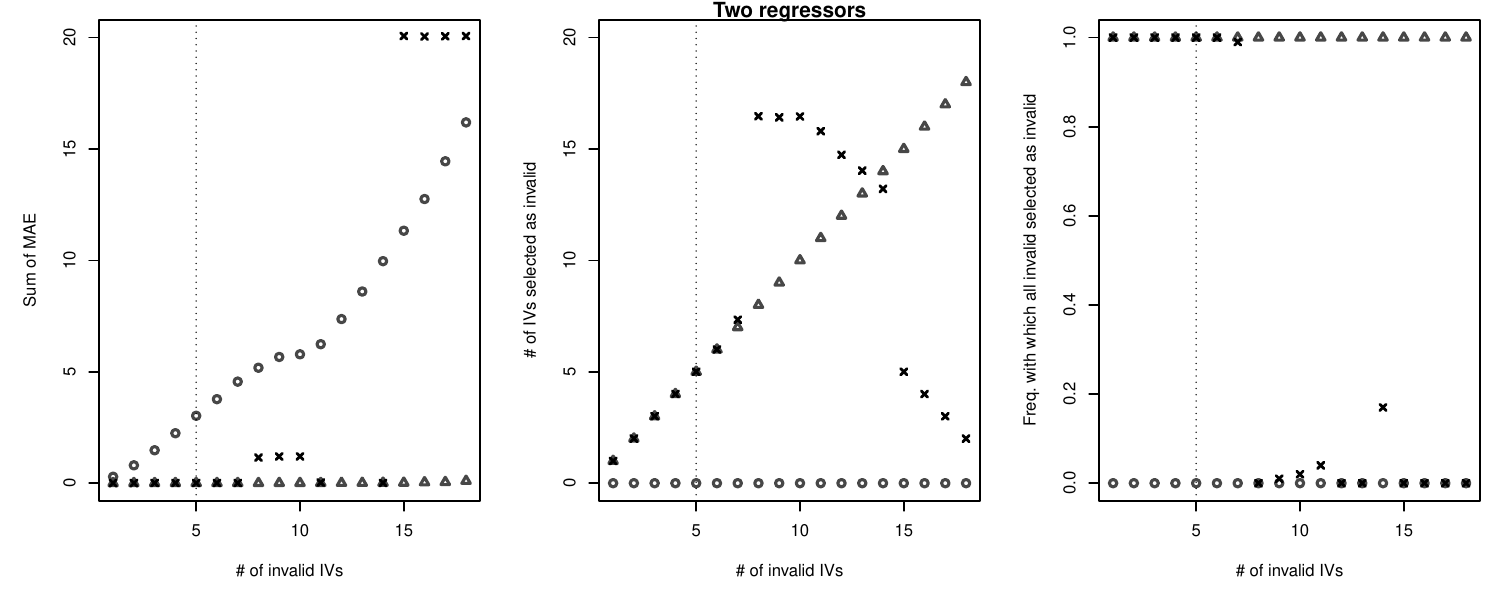}
			\includegraphics[scale=0.65, trim=0 0 0 0, clip]{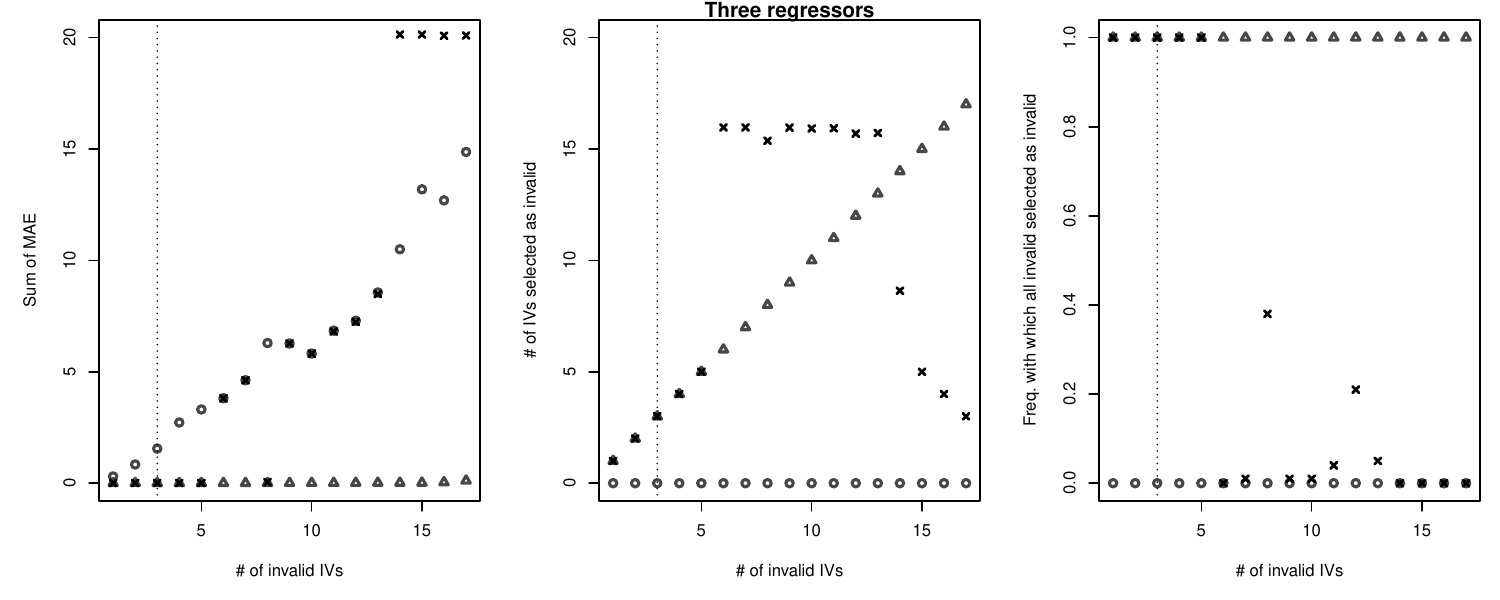}
			\caption{Simulation results multiple}
			\floatfoot{\textit{Note:} Performance of standard, oracle and post-adaptive Lasso 2SLS in Monte Carlo simulations as described in section \ref{sec:MCsim} using shares as IVs, with one, two and three endogenous regressors. 100 replications have been used for each number of invalid IVs. IVs chosen as invalid are included as controls. Horizontal axis: Number of invalid IVs. First column of graphs: median absolute deviation, second column: number of IVs chosen as invalid, third column: relative frequency with which all invalid IVs have been chosen as invalid. Circles: oracle 2SLS, triangles: standard 2SLS, crosses: post-AL 2SLS.\label{fig:SimulRes-Mult}}
		\end{center}
	\end{figure}

	The results can be found in figure \ref{fig:SimulRes-Mult}. As expected, the performance of the adaptive Lasso with the vector of marginal medians breaks down as soon as the majority rule is violated, i.e. when more than ten out of twenty IVs are invalid. 
	When there are two endogenous regressors the performance of the adaptive Lasso diverges from that of the oracle estimator when there are seven or more invalid IVs. The consistency of the vector of marginal medians is guaranteed only as long as five or less IVs are invalid. In this particular setting, however, the AL continues performing well when six IVs are invalid. This can be the case when some just-identified estimates that use invalid IVs end up above and some below the consistent estimates, making the qualified majority assumption stricter than needed. 
	When there are three endogenous regressors, again the AL continues performing as well as the oracle as long as five or less instruments are invalid.
	
	Overall, the results suggest that the extension of the adaptive Lasso also has oracle properties as long as sufficiently many instruments are valid, with the qualified majority condition becoming stricter with the number of invalid instruments.

	\newpage
	\begin{landscape}
		\begin{center}
			\section{Figures}	
			\begin{figure}[htb]
				\caption{Simulation results when majority valid\label{fig:All-maj}}
				\includegraphics[scale=0.45, trim=0 145 0 0, clip]{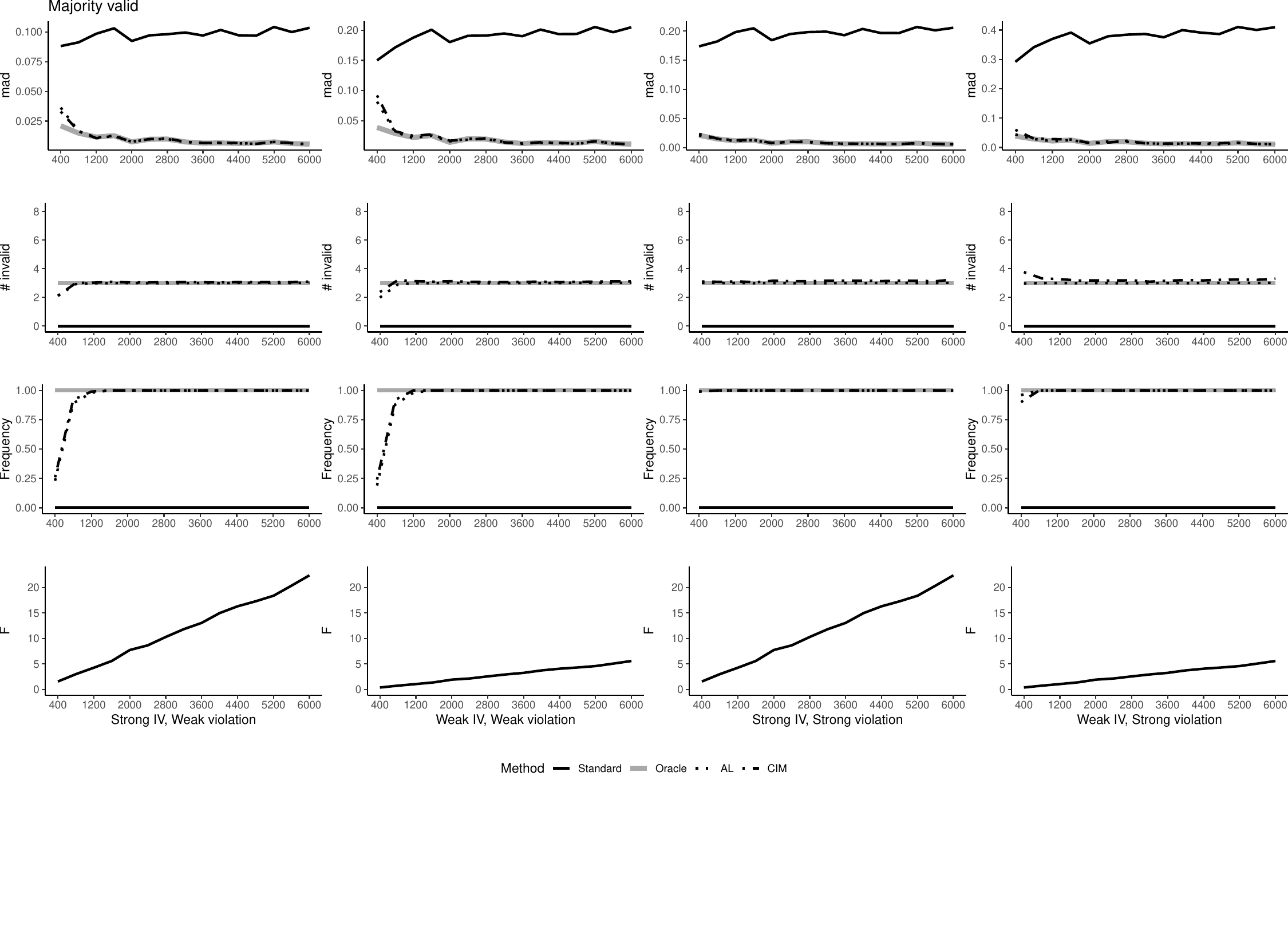}
				\floatfoot{\footnotesize \textit{Note:} Performance of 2SLS in Monte Carlo simulations as described in section \ref{sec:MCsim} using shares as IVs, adjusted with AL and confidence interval method. 100 MC replications for each sample size. IVs chosen as invalid are included as controls. Horizontal axis: Number of observations. First row: median absolute deviation, second row: number of IVs chosen as invalid, third row: relative frequency with which all invalid IVs have been chosen as invalid. Fourth row: F-statistic of oracle model. Grey line: oracle 2SLS, black line: standard 2SLS, dotted line: post-AL 2SLS, dashed-dotted line: post-Confidence Interval Method 2SLS. }
			\end{figure}
			\begin{figure}[htb]
				\caption{Simulation results when plurality valid\label{fig:All-plu}}
				\includegraphics[scale=0.45, trim=0 145 0 0, clip]{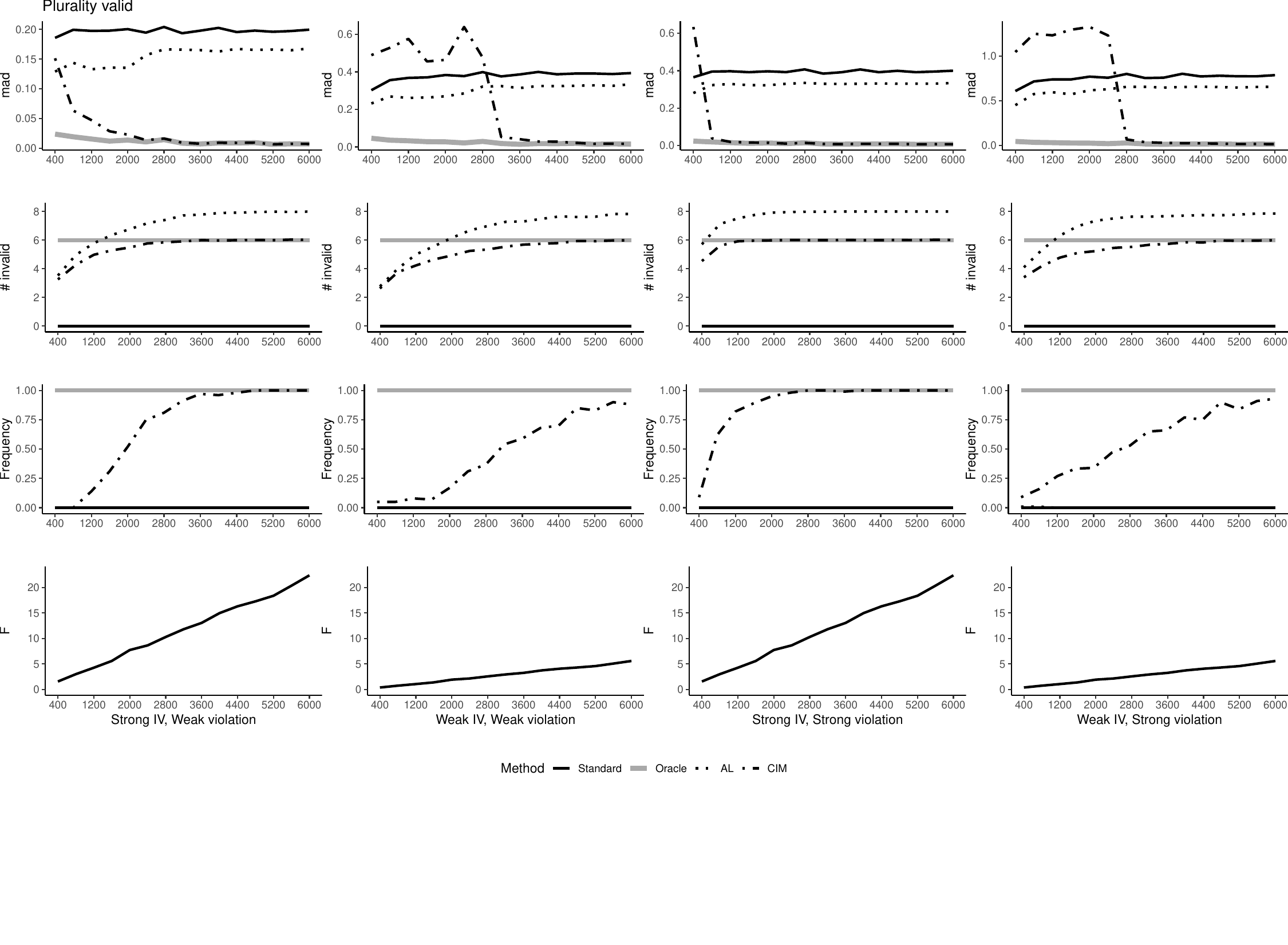}
				\floatfoot{\footnotesize \textit{Note:} Same as in figure \ref{fig:All-maj}, but with plurality valid.}
			\end{figure}
		\end{center}
	\end{landscape}
	
	\FloatBarrier
	\newgeometry{left=1.2in, right=1.2in, top=1.2in, bottom=1.2in}
	\section{Tables}
	
	\begin{center}
\begin{table}[h]
\caption[Justification of exclusion restriction in the literature]{Justification of exclusion restriction in the literature - Part 1 \label{tab:Quotes}}
\begin{tabularx}{\textwidth}{p{3cm} p{1.5cm} p{9.1cm} }
	Author & Journal & Citation\\
	\hline
	\citet{Amior2020Contribution} & Working Paper (WP) & \scriptsize{``The enclave instrument's validity depends on the exogeneity of the initial (origin-specific)
	migrant population shares (Goldsmith-Pinkham, Sorkin and Swift, 2018).''}\\
	
	\citet*{Edo2019Immigration} & Europ. Econ. Rev. (EER) & \scriptsize{``The identifying assumption is that the distribution of immigrants in 1968 is not correlated with voting [...]. This exclusion restriction means that, for instance, local economic shocks in 1968 are not correlated with voting more than 20 years later [...]. The assumption would be invalid if the initial distribution of immigrants is correlated with persistent
		local factors that influence future votes.''} \\
	
	\citet*{Bratti2018Effect} & Reg. Stud. & \scriptsize{``The main identifying assumption is
		that [...] the between-province
		variation within the same NUTS-2 region in the distribution of immigrants by different nationalities in 1995
		was approximately random with respect to provinces' future
		innovation prospects.''} \\
	
	\citet*{Aydemir2017Quasi} & EER & \scriptsize{`` The validity of our instrument requires that the ratio of earlier repatriates to non-repatriates
		across locations be unrelated to the change in unemployment rate from 1985 to 1990 in any
		way other than through its effect on the number of 1989 repatriates. [...]
		The key concern as to the validity of our instrument is that if earlier repatriates chose their
		locations based on economic circumstances, we could expect their location of residence in
		1985 to be related to the change in the economic conditions from 1985 to 1990 in that location.''}\\
	
	\citet*{Hunt2017Impact} & J. of Human Res. (JHR) & \scriptsize{``The instrument will be
		invalid if nonimmigration shocks to high school completion are correlated with 1940
		immigrant densities''.} \\

	\citet*{Foged2016Immigrants} & AEJ: Applied & \scriptsize{`` The plausibility of the exclusion restriction is predicated on the independence of the dispersal policy from labor demand conditions. '' Note: the dispersal policy determines the shares.} \\
	
	\citet*{Moreno-Galbis2016Effects} & EER & \scriptsize{``For our
		exclusion restriction to be valid, we require the natives' distribution within each educational group across occupations [...] to be independent from immigrants' labor supply shock.''} \\
	
	\citet*{Basso2015Association} & WP & \scriptsize{``This instrument is based on the idea that the distribution of foreign born of nationality $c$ in CZ $i$ in $t_0$ is uncorrelated with subsequent demand shifts and productivity
		changes in that CZ.''} \\
	
	\citet*{Bosetti2015Migration} & J. of Int. Econ. & \scriptsize{``The underlying exclusion restriction for this instrument is that the 1991 settlement of migrants by origin is not correlated with the economic situation after 1996. ''}\\
	
	\citet*{Cattaneo2015What} & JHR & \scriptsize{`` The assumption behind this instrument is that the distribution of immigrants of specific nationality across countries or occupations in 1991 is the result of historical settlements and past historical events. ''
		
		`` It should, therefore, be correlated with the share of foreign-born, but not with the region-sector specific demand shocks. ''}\\
	
	\citet*{Dustmann2015How} & J. of Labor Econ. (JoLE) & \scriptsize{``The idea
		is that immigrants tend to settle in areas in which other immigrants of the same country of origin have already settled earlier [...] but that these historical settlement patterns are not related to current demand-induced changes in local labor supply.'' 
		
		``Under the plausible assumption that current regional demand-induced labor market shocks are uncorrelated with past immigrant settlement patterns, this instrument leads to estimates that have a causal interpretation.''} \\
	
	\citet*{Kerr2015Skilled} & JoLE & \scriptsize{The authors mention a concern about the shift-share IV: 	``whether the initial distribution of country groups for skilled immigrants used in the interaction is correlated with something else that affects the measured outcomes.''}\\	
	
	\hline
\end{tabularx}
\begin{center}
(continued on next page)
\end{center}
\end{table}

\begin{table}
	\caption*{Justification of exclusion restriction in the literature - Part 2}
	\begin{tabularx}{\textwidth}{p{3cm} p{1.5cm} p{9.1cm} }
		Author & Journal & Citation\\
		\hline
	
\citet*{Orrenius2015Does} & JoLE & \scriptsize{``To be valid, the instrument requires assuming that the distribution of immigrants by country or region of origin across states 10 years ago is not correlated with shocks that affect the probability that natives in a state major in a STEM field 10 years later.''} \\
	
\citet*{Peri2015STEM} & JoLE & \scriptsize{The IV's ``validity is based, in large part, on the assumption that the 1980 employment share of foreign STEM workers varied across cities because of factors related to the persistent agglomeration of foreign communities in some localities. These historical differences [...] affected the change in the supply of foreign STEM workers but were unrelated to shocks affecting city-level native wage and employment growth. [...] For example, the initial distribution of foreign STEM may be correlated with persistent city factors that influenced future labor market outcomes, resulting in omitted variables bias.''} \\

\citet*{Damuri2014Immigration} & J. of the Europ. Econ. Assoc. (JEEA) & \scriptsize{``The underlying assumption is that while new immigrants tend to settle where existing immigrant communities already exist, in order to exploit ethnic networks and amenities, their historical presence is unrelated to current cell-specific changes in labor demand. [...] Current changes in labor demand have no correlation with the past presence of immigrants, which only affects the supply of labor and skills in that cell.''} \\

\citet*{Dustmann2013Effect} & Rev. of Econ. Stud. (REStud) & \scriptsize{``We instrument the change in this ratio using two alternative but closely related instruments: the 1991 ratio of immigrants to natives for each of these regions, from the Census of Population, interacted with	year dummies, and four period lags of the ratio of immigrants to natives in each region from the LFS.'' Note that shares are used as IVs directly.} \\

\citet*{Smith2012Impact} & JoLE & \scriptsize{``The exclusion restriction for this instrument requires that the composition of the immigrant population in t-1 [...] affects changes in native labor market outcomes only through its effect on changes in immigrant stocks.''} \\

\citet*{Cortes2011Low} & AEJ: Applied & \scriptsize{``The instrument will help in identifying the causal effect of immigration concentration on time use of native women as long as the following conditions hold: (1) The unobserved factors determining that more immigrants decided to locate in city $i$ versus city $i'$ (both cities in the same region) in 1970 are not correlated with changes in the relative economic opportunities for skilled women offered by the two cities during the 1980s and 1990s.''} \\

\citet*{Farre2011Immigration} & BE J. of Econ. Anal. \& Pol.& \scriptsize{``Our exogeneity assumption is that regional shocks to the demand for	female skilled labor between 1999 and 2008 are uncorrelated with immigrant location patterns prior to 1991.''}\\

\citet*{Dustmann2005Impact} & Economic J. & \scriptsize{``Pre-existing immigrant concentrations are unlikely to be correlated with current economic shocks if measured with a sufficient time lag, since 
existing concentrations are determined not by current economic conditions, but 
by historic settlement patterns of previous immigrants.''}\\

	\cite{Ottaviano2005Rethinking} & NBER WP & \scriptsize{``Since the instrument uses only the
	initial composition of foreign-born residents in a city and subsequent average immigration rates in the U.S. by
	nationality, it is not correlated with any city-specific factor that would affect actual immigration in the city
	during the decade. As a consequence it is by construction orthogonal to any city-specific shock to productivity,
	amenities and labor market conditions.''}\\

		\hline
	\end{tabularx}
\end{table}

\end{center}

\afterpage{
	\begin{landscape}
		\begin{center}
			\begin{table}[!htb]
				\caption{Impact of Immigration\label{tab:BP-mult}}
				{
\def\sym#1{\ifmmode^{#1}\else\(^{#1}\)\fi}
\begin{tabular*}{\textwidth}{@{\hskip\tabcolsep\extracolsep\fill}l*{9}{C{1.5cm} C{1.5cm} C{1.5cm} | C{1.5cm}  C{1.5cm} C{1.5cm} | C{1.5cm} C{1.5cm} C{1.5cm} C{1.5cm} }}
                    &\multicolumn{1}{c}{(1)}&\multicolumn{1}{c}{(2)}&\multicolumn{1}{c}{(3)}&\multicolumn{1}{c}{(4)}&\multicolumn{1}{c}{(5)}&\multicolumn{1}{c}{(6)}&\multicolumn{1}{c}{(7)}&\multicolumn{1}{c}{(8)}&\multicolumn{1}{c}{(9)}\\
                    &\multicolumn{1}{c}{Standard}&\multicolumn{1}{c}{AL (HS)}&\multicolumn{1}{c}{AL (AR)}&\multicolumn{1}{c}{Standard}&\multicolumn{1}{c}{AL (HS)}&\multicolumn{1}{c}{AL (AR)}&\multicolumn{1}{c}{Standard}&\multicolumn{1}{c}{AL (HS)}&\multicolumn{1}{c}{AL (AR)}\\
\toprule \multicolumn{1}{l}{DV:} & \multicolumn{3}{c}{Wages} & \multicolumn{3}{c}{High-skilled} & \multicolumn{3}{c}{Low-skilled} \\ \midrule SSIV $\Delta immi_t$&       5.194         &       5.194         &       3.329\sym{***}&       4.703         &       4.864         &       6.153         &       1.046         &       1.046         &      -3.011         \\
                    &     (6.727)         &     (6.727)         &     (0.934)         &     (6.469)         &     (6.941)         &     (4.689)         &     (0.823)         &     (0.823)         &     (10.24)         \\
\hspace{0.7cm} $\Delta immi_{t-10}$&      -2.083         &      -2.083         &      -1.297\sym{***}&      -1.778         &      -1.838         &      -2.620         &      -0.695\sym{*}  &      -0.695\sym{*}  &       0.709         \\
                    &     (2.583)         &     (2.583)         &     (0.340)         &     (2.485)         &     (2.658)         &     (2.708)         &     (0.324)         &     (0.324)         &     (3.689)         \\
J                   &       0.278         &       0.278         &       1.356         &       0.278         &       0.263         &       0.248         &       0.278         &       0.278         &       0.170         \\
\end{tabular*}
}
\hspace{-0.2cm}
				{
\def\sym#1{\ifmmode^{#1}\else\(^{#1}\)\fi}
\begin{tabular*}{\textwidth}{@{\hskip\tabcolsep\extracolsep\fill}l*{9}{C{1.5cm} C{1.5cm} C{1.5cm} | C{1.5cm}  C{1.5cm} C{1.5cm} | C{1.5cm} C{1.5cm} C{1.5cm} C{1.5cm} }}
\midrule TSLS $\Delta immi_t$&       1.126\sym{**} &       1.126\sym{**} &      -0.484         &       0.877\sym{*}  &       1.249\sym{***}&      -0.494\sym{+}  &       0.864\sym{**} &       0.864\sym{**} &      -0.278         \\
                    &     (0.371)         &     (0.371)         &     (0.399)         &     (0.343)         &     (0.367)         &     (0.297)         &     (0.273)         &     (0.273)         &     (0.183)         \\
\hspace{0.7cm} $\Delta immi_{t-10}$&      -0.439\sym{*}  &      -0.439\sym{*}  &       0.510\sym{+}  &      -0.249         &      -0.709\sym{**} &       0.411\sym{+}  &      -0.476\sym{***}&      -0.476\sym{***}&       0.222\sym{+}  \\
                    &     (0.180)         &     (0.180)         &     (0.288)         &     (0.173)         &     (0.250)         &     (0.216)         &     (0.124)         &     (0.124)         &     (0.126)         \\
J                   &       38.47         &       38.47         &       26.92         &       38.47         &       41.72         &       30.55         &       38.47         &       38.47         &       28.85         \\
\end{tabular*}
}

				{
\def\sym#1{\ifmmode^{#1}\else\(^{#1}\)\fi}
\begin{tabular*}{\textwidth}{@{\hskip\tabcolsep\extracolsep\fill}l*{9}{C{1.5cm} C{1.5cm} C{1.5cm} | C{1.5cm}  C{1.5cm} C{1.5cm} | C{1.5cm} C{1.5cm} C{1.5cm} C{1.5cm} }}
\midrule LIML $\Delta immi_t$&       5.984         &       5.984         &      -3.049         &       10.04         &       5.988         &      -2.845         &       3.855         &       3.855         &      -1.197\sym{+}  \\
                    &     (18.59)         &     (18.59)         &     (2.210)         &     (109.6)         &     (10.40)         &     (2.130)         &     (10.04)         &     (10.04)         &     (0.670)         \\
\hspace{0.7cm} $\Delta immi_{t-10}$&      -3.253         &      -3.253         &       2.310         &      -5.571         &      -4.137         &       1.966         &      -2.202         &      -2.202         &       0.899\sym{+}  \\
                    &     (11.01)         &     (11.01)         &     (1.582)         &     (64.22)         &     (7.492)         &     (1.450)         &     (5.946)         &     (5.946)         &     (0.498)         \\
J                   &       38.47         &       38.47         &       26.92         &       38.47         &       41.72         &       30.55         &       38.47         &       38.47         &       28.85         \\
\midrule \# inv     &           0         &           0         &          10         &           0         &           2         &          11         &           0         &           0         &           9         \\
Sign.               &           -         &         0.1         &     0.01302         &           -         &         0.1         &     0.01302         &           -         &         0.1         &     0.01302         \\
\bottomrule          \end{tabular*} }

				\vspace{-8px}
				\floatfoot{\footnotesize \textit{Note:} This table reports estimates of $\beta$ in equation \ref{eq:BP-dynamic}. The number of observations is $N= 2166$. Standard errors (in parentheses) are clustered by commuting zone. J denotes the Cragg-Donald test statistic. Observations are weighted by beginning-of-period population. Outcome variables are listed in the ``DV'' line. In the first row of each block separated by horizontal lines, results for contemporaneous immigration are reported and in the second line, results for lagged immigration are reported. In the last two rows, the number of countries chosen as invalid and the thresholds used in the HS procedure are reported. In columns 1, 4 and 7 all shares are assumed to be valid. Column heads of columns 2, 3, 5, 6, 8 and 9 denote which method has been used for selection.}
			\end{table}
		\end{center}
	\end{landscape}
}
\FloatBarrier

	\afterpage{
		\begin{landscape}
			\begin{center}
				\begin{table}[!htb]
					\caption{Impact of Immigration\label{tab:BP-Shifts2}}
					{
\def\sym#1{\ifmmode^{#1}\else\(^{#1}\)\fi}
\begin{tabular*}{\textwidth + 4\tabcolsep}{@{\hskip\tabcolsep\extracolsep\fill}l*{9}{C{2cm} C{2cm} C{2cm}  | C{2cm} C{2cm} C{2cm} | C{2cm} C{2cm} C{2cm} C{2cm} }}
                    &\multicolumn{1}{c}{(1)}&\multicolumn{1}{c}{(2)}&\multicolumn{1}{c}{(3)}&\multicolumn{1}{c}{(4)}&\multicolumn{1}{c}{(5)}&\multicolumn{1}{c}{(6)}&\multicolumn{1}{c}{(7)}&\multicolumn{1}{c}{(8)}&\multicolumn{1}{c}{(9)}\\
                    &\multicolumn{1}{c}{Standard}&\multicolumn{1}{c}{AL (HS)}&\multicolumn{1}{c}{AL (AR)}&\multicolumn{1}{c}{Standard}&\multicolumn{1}{c}{AL (HS)}&\multicolumn{1}{c}{AL (AR)}&\multicolumn{1}{c}{Standard}&\multicolumn{1}{c}{AL (HS)}&\multicolumn{1}{c}{AL (AR)}\\
\toprule \multicolumn{1}{l}{DV:} & \multicolumn{3}{c}{Wages} & \multicolumn{3}{c}{High-skilled} & \multicolumn{3}{c}{Low-skilled} \\ \midrule 2SLS $\Delta immi_t$&       0.726\sym{***}&       0.290         &      -0.463         &       0.654\sym{***}&       0.583\sym{***}&      -0.341         &       0.626\sym{***}&       0.226         &      -0.147         \\
                    &     (0.151)         &     (0.181)         &     (0.555)         &     (0.159)         &    (0.0976)         &     (0.254)         &     (0.117)         &     (0.279)         &     (0.508)         \\
\hspace{0.7cm} $\Delta immi_{t-10}$&      -0.291\sym{**} &      -0.378\sym{**} &     -0.0620         &      -0.168\sym{*}  &      -0.438\sym{***}&      -0.139         &      -0.482\sym{***}&      -0.390\sym{*}  &      -0.115         \\
                    &    (0.0889)         &     (0.117)         &     (0.324)         &    (0.0670)         &    (0.0794)         &     (0.158)         &    (0.0735)         &     (0.193)         &     (0.266)         \\
J                   &       15.52         &       7.119         &       5.834         &       15.52         &       14.65         &       4.865         &       15.52         &       7.119         &       5.746         \\
\end{tabular*}
}
 \hspace{0.2cm}{
\def\sym#1{\ifmmode^{#1}\else\(^{#1}\)\fi}
\begin{tabular*}{\textwidth + 4\tabcolsep}{@{\hskip\tabcolsep\extracolsep\fill}l*{9}{C{2cm} C{2cm} C{2cm}  | C{2cm} C{2cm} C{2cm} | C{2cm} C{2cm} C{2cm} C{2cm} }}
\midrule LIML $\Delta immi_t$&       0.954         &       0.154         &      -0.488         &       0.822         &       0.569\sym{***}&      -0.376         &       0.762\sym{*}  &       0.180         &      -0.150         \\
                    &     (0.664)         &     (0.266)         &     (0.573)         &     (0.644)         &     (0.101)         &     (0.272)         &     (0.325)         &     (0.346)         &     (0.511)         \\
\hspace{0.7cm} $\Delta immi_{t-10}$&      -0.344\sym{***}&      -0.367\sym{*}  &     -0.0552         &      -0.168         &      -0.500\sym{***}&      -0.135         &      -0.551\sym{***}&      -0.389\sym{+}  &      -0.114         \\
                    &    (0.0942)         &     (0.186)         &     (0.333)         &     (0.106)         &     (0.108)         &     (0.169)         &    (0.0551)         &     (0.231)         &     (0.267)         \\
J                   &       15.52         &       7.119         &       5.834         &       15.52         &       14.65         &       4.865         &       15.52         &       7.119         &       5.746         \\
\midrule \# inv     &           0         &           3         &           8         &           0         &           2         &           7         &           0         &           3         &           7         \\
Sign.               &           -         &         0.1         &     0.01302         &           -         &         0.1         &     0.01302         &           -         &         0.1         &     0.01302         \\
\bottomrule          \end{tabular*} }

					\vspace{-8px}
					\floatfoot{\footnotesize \textit{Note:} This table reports estimates of $\beta$ in equation \ref{eq:BP-dynamic}. The number of observations is $N= 2166$. Standard errors (in parentheses) are clustered by commuting zone. J denotes the Cragg-Donald test statistic. Observations are weighted by beginning-of-period population. Outcome variables are listed in the ``DV'' line. In the first row of each block separated by horizontal lines, results for contemporaneous immigration are reported and in the second line, results for lagged immigration are reported. In the last two rows, the number of countries chosen as invalid and the thresholds used in the HS procedure are reported. In columns 1, 4 and 7 all shares are assumed to be valid. Column heads of columns 2, 3, 5, 6, 8 and 9 denote which method has been used for selection.}
				\end{table}
			\end{center}
		\end{landscape}
	}
	
	\begin{table}[!htb]
		\caption{Shocks selected as invalid}\label{tab:ExclShocksBP}
		\begin{scriptsize}
\begin{tabular}{*{14}{c}}
DV & Method & SL & \rot{Migration} & \rot{Battle-related deaths} & \rot{Onesided violence} & \rot{Nonstate violence} & \rot{Population} & \rot{FH Civil Liberties} & \rot{FH Political} & \rot{FH Status} & \rot{Polity} & \rot{Press Freedom Status} & \rot{Press Freedom Score}\\
\toprule
dlweekly & AL (HS) & 0.1 & - & x & - & - & - & - & x & - & - & - & x \\
dlweekly & AL (AR) & 0.01302 & - & x & x & x & - & x & x & x & - & x & x\\
dlweekly\_hskill & AL (HS) & 0.1 & - & x & - & - & - & - & x & - & - & - & - \\
dlweekly\_hskill & AL (AR) & 0.01302 & - & x & x & x & - & x & x & x & - & - & x\\
dlweekly\_lskill & AL (HS) & 0.1 & - & x & - & - & - & - & x & - & - & - & x \\
dlweekly\_lskill & AL (AR) & 0.01302 & - & x & x & - & x & x & x & - & - & x & x\\
 \bottomrule
 \multicolumn{14}{c}{%
 	\begin{minipage}{\textwidth - 18\tabcolsep}%
 		\vspace{2px}
\textit{Note:} This table reports the countries chosen as invalid in tables \ref{tab:BP-SSIV} and \ref{tab:BP-mult-SSIV} for the reanalysis of \citet*{Basso2015Association}. The left columns display the method and the outcome variable used. x denotes a shock selected as invalid.
 	\end{minipage}%
 }
\end{tabular}
\end{scriptsize}
	\end{table}	
	
	\begin{table}[!htbp]
		\begin{footnotesize}
	\begin{tabularx}{\textwidth-2\tabcolsep}{ccX}
	Analysis & Table, Column & Countries / Excluded SIC codes\\
	\toprule
	\multicolumn{3}{c}{China Shock}\\
	\midrule
	AL & \ref{tab:ADH}, 3 & Broadwoven Fabric Mills, Manmade Fiber and Silk (2221)\\
	
	CIM & \ref{tab:ADH},5 & Poultry Slaughtering and Processing (2015), (2221), Aluminum Foundries (3365), Ordnance and Accessories, Nec (3489), Industrial Patterns (3543), Household Audio and Video Equipment (3651), Electronic Components, Nec (3679), Measuring and Controlling Devices, Nec (3829) \\
	
	AL AR	& \ref{tab:ADH}, 6 & Food (20):	8,
	Textile Mill Products (22):	4,
	Apparel \& other (23):	4,
	Lumber \& Wood (24):	1,
	Furniture (25):	2,
	Paper (26):	2,
	Chemicals and allied (28):	4,
	Leather (31):	2,
	Stone, Clay, Glass, and Concrete (32):	3,
	Primary Metal Industries (33):	2,
	Fabricated Metal Prdcts, Except Machinery \& Transport (34):	4,
	Industrial and Commercial Machinery and Computer Equipment (35):	10,
	Electronic, Electrical Eqpmnt \& Cmpnts, Excpt Computer Eqpmnt (36):	8,
	Transportation Equipment (37):	5,
	Mesr/Anlyz/Cntrl Instrmnts; Photo/Med/Opt Gds (38):	2,
	Miscellaneous Manufacturing Industries (39):	2
	 \\
	
	CIM AR & \ref{tab:ADH}, 7 & Food (20): 13,
	Tobacco (21): 1,
	Textile Mill Products (22): 5,
	Apparel \& other (23): 13,
	Lumber \& Wood (24): 6,
	Furniture (25): 5,
	Paper (26): 3,	
	Chemicals and allied (28): 8,
	Petroleum Refining and Related (29): 2,
	Rubber (30): 3,
	Leather (31): 3,
	Stone, Clay, Glass, and Concrete (32): 6,
	Primary Metal Industries (33): 8,
	Fabricated Metal Prdcts, Except Machinery \& Transport (34): 6,
	Industrial and Commercial Machinery and Computer Equipment (35): 16,
	Electronic, Electrical Eqpmnt \& Cmpnts, Excpt Computer Eqpmnt (36): 12,
	Transportation Equipment (37): 6,
	Mesr/Anlyz/Cntrl Instrmnts; Photo/Med/Opt Gds (38): 8,
	Miscellaneous Manufacturing Industries (39): 4
	\\ 
	\bottomrule
\end{tabularx}
\end{footnotesize}
		\caption{Industries chosen as invalid \label{tab:ExclCountries}}
		\floatfoot{\footnotesize \textit{Note:} This table reports industries chosen as invalid in table \ref{tab:ADH} for the reanalysis of \citet*{Autor2013china}. The left column displays the method and the outcome variable used. The middle column reports table and column number, in which the analysis can be found. The right column displays the industries selected as invalid. For AL AR and CIM AR the number of 4-digit industries in each 2-digit industry are reported, for brevity.}
	\end{table}
	
		\newpage
	\FloatBarrier	
	\section{Documentation for ado-files}
	
	The following subsection provides the documentation for the \verb+ssada+ - and \verb+sscim+ - programs in Stata. 
	\textbf{Preliminaries:} Save \verb+ssada+ and \verb+sscim+ to your personal ado-directory.
	
	\subsection{Adaptive Lasso shift-share}
	The Stata implementation of AL shift-share is called \verb+ssada+. The code is a variation of \verb+sivreg+ \citep*{Farbmacher2017SIVREG} and shares its syntax. The differences are that in \verb+ssada+ analytical weights are allowed, the standard errors to be reported in the post-AL regression can be chosen and locals containing valid and invalid IVs are returned. \verb+moremata+ is required.
	
	\paragraph{Syntax}
	\begin{verbatim}
		ssada depvar indepvars [if] [in] [aw], ///
		endog(varlist) exog(varlist) id(string) [options]
	\end{verbatim}
	
	\paragraph{Options}
	$\quad$
	\linebreak
	\textbf{Required:}\\
	\begin{tabularx}{\textwidth}{sX}
		\verb+endog+ & Endogenous variable\\
		\verb+exog+ & Exogenous controls as well as potentially endogenous shares used for construction of the shift-share IV.  The shares should have the following naming: e.g. \textbf{stub}1, \textbf{stub}2, \textbf{stub}3, ...\\
		\verb+id+ & String denoting variables by which observations are identified\\
	\end{tabularx}
	$\quad$
	\linebreak
	\textbf{Optional:}\\
	\begin{tabularx}{\textwidth}{sX}
		\verb+aw+ & Analytical weights (\verb+aweight+) are allowed\\
		\verb+vce+ & Specifies the type of standard error reported. Same as in standard \verb+vce-option+. Default is \verb+robust+ \\
		\verb+c+ & \verb+real+ specifying the significance level as $c/ln(n)$ for the Andrews-Hansen stopping rule. Default is 0.1 \\
	\end{tabularx}
	
	\paragraph{Stored results}
	
	\verb+ssada+ stores the results of the last post-AL \verb+ivregress+-command in \textbf{e()}. Moreover, the following macros are returned:
	
	\begin{tabularx}{\textwidth}{sX}
		\textbf{e(wv)} & A local containing the varnames of variables chosen as valid by the AL algorithm\\
		\textbf{e(wi)} & A local containing the varnames of variables chosen as invalid by the AL algorithm\\
	\end{tabularx}
	
	\newpage
	
	\subsection{Confidence Interval Method shift-share}
	
	The setup of \verb+sscim+ is adopted from \verb+ssada+ (a large  part of lines 1-150).
	
	\paragraph{Syntax}
	\begin{verbatim}
		sscim depvar indepvars [if] [in] [aw], ///
		endog(varlist) exog(varlist) ssstub(string) [options]
	\end{verbatim}
	
	\paragraph{Options}
	$\quad$
	\linebreak
	\textbf{Required:}\\
	\begin{tabularx}{\textwidth}{sX}
		\verb+endog+ & Endogenous variable\\
		\verb+exog+ & Exogenous controls as well as potentially endogenous shares used for construction of the shift-share IV\\
		\verb+ssstub+ & Stub of shares. The shares should have the following naming: e.g. \textbf{stub}1, \textbf{stub}2, \textbf{stub}3, ...\\
	\end{tabularx}
	$\quad$
	\linebreak
	\textbf{Optional:}\\
	\begin{tabularx}{\textwidth}{sX}
		\verb+aw+ & Analytical weights (\verb+aweight+) are allowed\\
		\verb+vce+ & Specifies the type of standard error reported. As in \verb+vce()+-option in \verb+ivreg2+, e.g. \verb+vce("cluster(state)")+. Default is \verb+"robust"+ \\
		\verb+c+ & \verb+real+ specifying the significance level as $c/ln(n)$ for the Andrews-Hansen stopping rule. Default is 0.1 \\
		\verb+psif+ & Specifies initial critical value with which confidence intervals are calculated, according to $\psi=$\verb+psif+$\times\sqrt{2.01^2*ln(N)}$. Set this larger than one if in the beginning already more than one IV are chosen as invalid. Default is 1.\\
	\end{tabularx}
	
	\paragraph{Stored results}
	
	\verb+sscim+ stores the results of the last post-CIM \verb+ivregress+-command in \textbf{e()} and the following macros:
	
	\begin{tabularx}{\textwidth}{sX}
		\textbf{e(wv)} & A local containing the varnames of variables chosen as valid by the confidence interval method\\
		\textbf{e(wi)} & A local containing the varnames of variables chosen as invalid by the confidence interval method\\
	\end{tabularx}

	\paragraph{Post-estimation}
	$\quad$
	\linebreak
	For both \verb+ssada+ and \verb+sscim+, the same post-estimation results as for \verb+ivregress+ apply.

\end{appendices}
\newpage

\end{document}